\documentclass[12pt]{article}
\usepackage{amsmath,amsfonts, epsfig}
\usepackage{fullpage,feynmf}
\usepackage{amssymb}
\usepackage{amscd}
\advance \textheight by \topskip
\makeatletter

\@addtoreset{equation}{section}
 \def\theequation{\thesection.\arabic{equation}}
\makeatother

\newtheorem{theorem}{Theorem}[section]

\newtheorem{definition}[theorem]{Definition}

\newtheorem{lemma}[theorem]{Lemma}

\newtheorem{proposition}[theorem]{Proposition}
\newtheorem{remark}[theorem]{Remark}

\def\One{\mathbb{I}}
\def\KK{{\mathbb K}}
\def\NN{{\mathbb N}}

\def\NN{{\mathbb N}}

\newcommand{\beqa}{\begin{eqnarray}}
\newcommand{\eeqa}{\end{eqnarray}}
\newcommand{\noi}{\noindent}
\newcommand{\e}{\varepsilon}

\newcommand{\ch}{{\cal H}}

\def\>{\rangle}
\def\<{\langle}

\begin{document}

\title{
\begin{flushright}
{\small CPHT-RR048.0609}\\[-0.4cm]
\end{flushright}
 \vskip .5in
{\bf Combinatorial Dyson-Schwinger equations in noncommutative field theory
}  
\author{ 
{\sf   Adrian Tanasa${}^{a,b}$\thanks{e-mail:
kreimer@ihes.fr}
 and
{\sf Dirk Kreimer}${}^{c,d}$\thanks{e-mail:
adrian.tanasa@ens-lyon.org}}\\
{\small ${}^{a}${\it Centre de Physique Th\'eorique, CNRS, UMR 7644,}} \\
{\small {\it \'Ecole Polytechnique, 91128 Palaiseau, France}}  \\ 
{\small ${}^{b}${\it Institutul de Fizic\u a \c si Inginerie Nuclear\u a Horia Hulubei,}} \\
{\small {\it P. O. Box MG-6, 077125 M\u agurele, Rom\^ania}}\\
{\small ${}^{c}${\it Institut des Hautes \'Etudes Scientifiques, CNRS,}}\\
{\small {\it Le Bois-Marie 35, route de Chartres, 91440 Bures sur Yvette, France}}\\
 {\small ${}^{d}${\it Center for Mathematical Physics, Boston University,}}\\
{\small {\it 111 Cummington
Street, Boston MA, 02215, USA}}\\
}}
\maketitle
\vskip-1.5cm

\vspace{2truecm}

\begin{abstract}
\noindent
We introduce here the Hopf algebra structure describing the noncommutative renormalization of a recently introduced translation-invariant
model on Moyal space. We define Hochschild one-cocyles $B_+^\gamma$ which allows us to write down the combinatorial Dyson-Schwinger equations 
for noncommutative quantum field theory. One- and two-loops examples are explicitly worked out.  
\end{abstract}


\newpage

\section{Introduction}
\renewcommand{\theequation}{\thesection.\arabic{equation}}    
\setcounter{equation}{0}

Hochschild cohomology was shown, in the context of {\it commutative} quantum field theory (QFT), 
to play an important r\^ole in the understanding of different perturbative and non-perturbative issues \cite{bk,YK}. 
Using suitable Hochschild  $1-$cochains $B_+^\gamma$, one can thus write down the combinatorial Dyson-Schwinger equation, 
extending perturbative to non-perturbative physics.

On the other hand, noncommutative geometry is an interesting framework for both mathematics and theoretical physics 
(see for example \cite{book1, book2}).
{\it Noncommutative} quantum field theory (NCQFT) on Moyal space has recently gained attention through 
the proposition of several models which were proved to be perturbatively renormalizable. 
Thus, despite the ultraviolet/infrared mixing problem \cite{min} (a new type of non-local divergence which appears when 
implementing QFT on the Moyal space), several models are now known to be renormalizable. 
A first such model is the Grosse-Wulkenhaar model \cite{GW2}, which however explicitly breaks the translation-invariance of QFT. 
Recently, a translation-invariant model was proposed in \cite{noi}; this new model was also proved to be renormalizable at any order in 
perturbation theory \cite{noi}.

The Hopf algebra structure of perturbative renormalization was implemented for the Grosse-Wulkenhaar model in \cite{hopf}. 
In this paper we first repeat this for the translation-invariant model \cite{noi}. This task is more involved because of a more 
complicated power counting mechanism. 
We then go further and introduce  Hochschild $1-$cocyles $B_+^\gamma$ adapted for this noncommutative framework. 
This allows us to write down the combinatorial Dyson-Schwinger equations for both these types of noncommutative models. 
Nothing here involves mathematical sophistication beyond what was necessary for commutative field theory. But finer technical details have to be clarified, and are done so here 
by explicit example.

\medskip

This paper is structured as follows. The next section recalls the definition of the Moyal space and lists the field theoretical models known so 
far to be renormalizable in this noncommutative frame. 
Particularly, we recall the translation-invariant model introduced in \cite{noi}. 
In the third section we introduce the Hopf algebra structure which describes the renormalization of this noncommutative model. 
The pre-Lie and Lie algebra structures associated to graphs are also presented. 
The fourth section analyzes in detail several differences (from a diagrammatic point of view) which appear 
when one uses the ribbon graph representation of NCQFT instead of the usual Feynman graphs of commutative QFT. 
In section $5$ we introduce  Hochschild $1-$cocycles $B_+^\gamma$ which allow to write down the combinatorial Dyson-Schwinger equations in NCQFT. 
We give here the corresponding theorems; note that these results hold for all renormalizable noncommutative models listed in section $2$. 
Finally, in the last section we completely work out as the crucial part of this paper the one- and two-loop implementations of the theorems 
of section $5$.

\section{Scalar field theory on the Moyal space and renormalizability} \label{mixing}
\renewcommand{\theequation}{\thesection.\arabic{equation}}   
\setcounter{equation}{0}
\label{rappel}

In this section we briefly recall the definition of the Moyal space; 
we then list the field theoretical models (translation-invariant or not) known so far to be renormalizable on it.

The noncommutative Moyal space is given by
\beqa
[x^\mu, x^\nu]_\star=\imath \Theta^{\mu \nu},
\eeqa
where the noncommutative matrix $\Theta$ writes 
\beqa
\label{theta}
\Theta=
\begin{pmatrix}
   0 &\theta & 0 & 0\\   
   -\theta & 0  & 0 & 0\\
   0&0 & 0 & \theta\\  
   0& 0 & -\theta & 0  
  \end{pmatrix}.  
\eeqa
Note that by $\star$ we denote the Moyal-Weyl product.

\subsection{ The ``naive''  $\phi^{\star 4}$ model; Feynman graphs (planarity and non-planarity) - ribbon graphs}

In order to obtain field theory on this space, the first thing that comes to mind is to 
 replace the ordinary commutative local product of fields   
 by the Moyal-Weyl product
\beqa
\label{act-normala}
S[\phi]=\int d^4 x \left(\frac 12 \partial_\mu \phi \star \partial^\mu \phi +\frac
12 \mu^2 \phi\star \phi  + \frac{\lambda}{4!} \phi \star \phi \star \phi \star \phi\right).
\eeqa
Note that an Euclidean metric is used. 

In momentum space the action (\ref{act-normala}) writes
\beqa
\label{act-normala-p}
S[\phi]=\int d^4 p \left(\frac 12 p_\mu \phi  p^\mu \phi +\frac
12 \mu^2 \phi  \phi  + \frac{\lambda}{4!} \phi \star \phi \star \phi \star \phi\right).
\eeqa

An important consequence of the utilization of the non-local product $\star$ 
is that the interaction part does not longer preserves the invariance under permutation of incoming (outgoing) fields. 
This invariance is restricted to a cyclical permutation.

 Furthermore, there exits a basis - the matrix base - of the Moyal algebra for which the Moyal-Weyl product becomes an ordinary matrix product. 
For these reasons, an appropriate way to draw the associated Feynman graphs is to use ribbon graphs, 
that is to use ribbons instead of lines for the propagators.

\begin{figure}
\centerline{\epsfig{figure=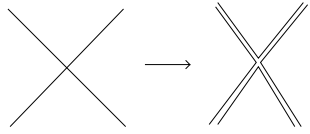,width=4cm} }
\caption{The local vertex is replaced in NCQFT with a non-local, Moyal vertex.}
\label{vertecsi}
\end{figure}

Thus, the ``usual'' commutative $\phi^4$ vertex becomes some ribbon $\phi^{\star\, 4}$ vertex as shown in Fig. \ref{vertecsi}.

This has important consequences on  the definition of the pre-Lie structure of insertions of such ribbon graphs (see subsection \ref{preLie}).
Let us first give an important topological definition on these ribbon graphs:

\begin{definition}
A planar graph is called { regular} if it has a single face (the external one) broken by external edges. 
\end{definition}
While this definition takes recourse to the parlance in noncommutative field theory, it just takes into account standard combinatorial facts:
Let us call a graph a genus-$n$ graph if it can be drawn without self-intersection on all surfaces with genus $\geq n$. Then, a planar graph is genus-0. 

\begin{figure}
\centerline{\epsfig{figure=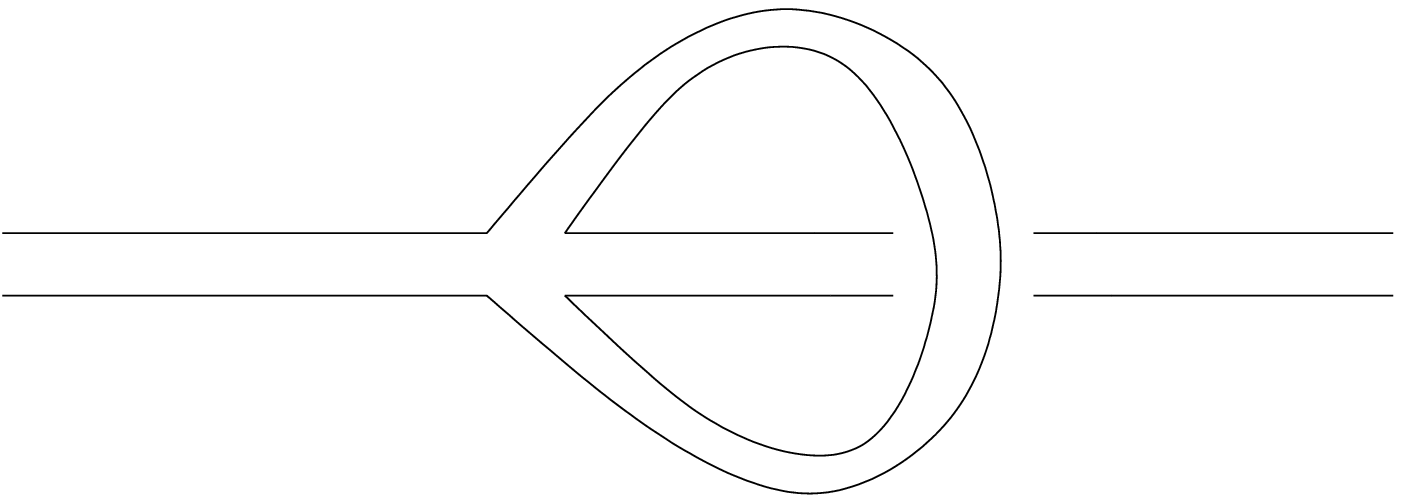,width=4cm} }
\caption{The ``non-planar'' tadpole. This graph is planar irregular (it has two broken faces)}
\label{nptad}
\end{figure}

Now, a genus-$n$ graph can be drawn on a surface of genus $n$ without self-intersections. Its internal edges and vertices hence allow for a unique labeling
of faces on this genus $n$ surface. Let $k$ be the number of such faces which contain an external edge of the graph. A planar regular graph is a genus-0
graph with $k=1$. If $k>1$, we call the graph irregular. An example of a $2-$point planar irregular graph is the tadpole of Fig. \ref{nptad}.

\subsection{The Grosse-Wulkenhaar-like models}
\label{GW}
The Grosse-Wulkenhaar model is a scalar quantum field theory on the four-dimensional Moyal space. Its action \cite{GW2} is given by
\begin{equation}\label{action}
S[\phi] = \int d^4x \Big( -\frac{1}{2} \phi(-\Delta)\phi + \frac{\Omega^2}{2}\tilde x^{2}\phi^{2}+ \frac{1}{2} m^2\,\phi^{2}
+ \frac{\lambda}{4!} \phi \star \phi \star \phi \star
\phi\Big)(x)
\end{equation}
with $\tilde x _\mu=2(\Theta^{-1}x)_{\mu}$. 
This model has been shown renormalizable to all orders of perturbation, using different field theoretical or algebraic methods.

Several field theoretical properties have been adapted to this type of noncommutative models 
(see \cite{dimreg}, \cite{param2} , \cite{param}, \cite{param'}, \cite{mellin}, \cite{goldstone} and references within). 
Some algebraic geometrical properties of the parametric representation of the Grosse-Wulkenhaar model have been proved in \cite{marcolli}.

Let us now give an elementary definition for what a primitive element in a Hopf algebra of graphs should mean for a physicist:
\begin{definition}
A primitive divergent graph of a quantum field theoretical model is a graph whose Feynman amplitude is divergent but which does 
not contain any subgraph for whom the Feynman amplitude is also divergent.
\end{definition}

For the Grosse-Wulkenhaar-like models, the primitive divergent graphs are the planar 
regular $2-$ and $4-$point graphs. A  Hopf algebra structure adapted for this noncommutative 
renormalization was defined in \cite{hopf}. 
The main idea is that the notion of locality, crucial in commutative field theory, 
is replaced by a new one, of ``Moyality'', stating that the counterterms will have the same, non-local ``Moyal'' form, 
as the terms in the original actions. For a more detailed discussion on this aspect the interested reader is referred  to \cite{beta-GMRT}.

\subsection{A translation-invariant scalar model}
Note that the Grosse-Wulkenhaar model (\ref{action}) is manifestly not translation-invariant. 
In order to preserve the translation-invariance, one possibility is to modify the propagation in a different way 
\cite{noi}
\beqa
\label{revolutie}
S[\phi]=\int d^4 p (\frac 12 p_{\mu} \phi  p^\mu \phi  +\frac
12 m^2  \phi  \phi   
+ \frac 12 a  \frac{1}{\theta^2 p^2} \phi  \phi  
+ V^\star [\phi] ),
\eeqa
where   $a$ some dimensionless parameter and $V^\star[\phi]$ is the corresponding potential in momentum space.
The propagator is
\beqa
\label{propa-rev}
\frac{1}{p^2+m^2+\frac{a}{\theta^2 p^2}}.
\eeqa
One further chooses $ a \ge 0 \, $
such that the propagator (\ref{propa-rev}) is positively defined.
In \cite{noi}, this model was proved to be renormalizable at any order in perturbation theory. 
Furthermore, its renormalization group flows \cite{beta-GMRT} and parametric representation \cite{param-GMRT} were implemented; 
a mechanism for taking the commutative limit has been proposed \cite{limita} (for a review on all these developments, 
refer to \cite{review-io}). Also, a propagator (\ref{propa-rev}) like above has appeared in recent work on non-abelian gauge theory 
in the context of the Gribov-Zwanziger result \cite{Gracey}. A connection between those result and a suitable noncommutative model 
is unknown though at the time of writing.

\section{Hopf algebra for the noncommutative model (\ref{revolutie}). Planar irregular graphs}
\renewcommand{\theequation}{\thesection.\arabic{equation}}   
\setcounter{equation}{0}

\subsection{Considerations on its primitive divergent graphs}

 As proved in \cite{noi}, the primitive divergent graphs of the translation-invariant model (\ref{revolutie}) are again 
the $2-$ and $4-$ point graphs. 
However, a more thorough discussion is requested here. 
In the case of the planar regular graphs, these $2-$ and $4-$point graphs will lead to the renormalization of the mass, 
field strength and coupling constant, just like in the case of the commutative $\phi^4$ model. 
A more tricky situation appears for the planar irregular graphs. 
The $4-$point function graphs are proved to be convergent. 
The $2-$point function graphs (which are the ones encoding the UV/IR mixing) are again {\it convergent}.
 Nevertheless, when going in the UV regime of their internal momenta, they lead a priori to an $1/p^2$ contribution to the 
respective Feynman amplitude ($p$ being the external momenta of the respective $2-$point graph). 
However, the modification of the propagation given in (\ref{revolutie}) will insure the renormalizability of the model. 
These $2-$ point planar irregular graphs will just lead to a {\it finite} renormalization for the coefficient $a$ in the action (\ref{revolutie}). 

When inserting these $2-$point graphs into bigger graphs, 
these latter become non-planar (for example, when inserting the tadpole of Fig \ref{nptad} in any planar ribbon graph, 
the resulting ribbon graph is non-planar). In the case of the ``naive'' model (\ref{act-normala}), 
the Feynman amplitudes of these graphs is UV convergent but IR divergent (because of the UV/IR mixing). 
In the case of the model (\ref{revolutie}), these non-planar graphs are also convergent in IR regime, 
thanks to the $1/p^2$ terms in the propagator (see again the proofs of \cite{noi} or \cite{param-GMRT}).

Let us conclude by stating that, 
for the reasons explained above, the primitive divergent graphs of the model (\ref{revolutie}) are taken to be the 
planar regular $2-$ and $4-$point graphs. 
See also subsection \ref{discard} below.

\subsection{Insertions of graphs; the pre-Lie and Lie structures}
\label{preLie}

In this subsection we explain the operation of insertion of graphs and the difficulties encountered when doing this for the ribbon graphs of NCQFT. These difficulties come from the fact that one has to deal with a non-local vertex with restricted symmetry (see Fig. \ref{vertecsi}). 
This insertion operations allows us to define the pre-Lie structure of Feynman ribbon graphs.

\begin{definition}
The residue {\bf res} of a ribbon graph is the graph obtained by shrinking all its internal edges.
\end{definition}
Note that any 4-point ribbon vertex can be assumed to lie in an infinitesimal  small disc with its edges giving a cyclic labeling of four points on the boundary 
of that disc. Furthermore, for a regular graph, its external edges define such a cyclic ordering on the distinguished face containing the external edges.

Let the set of residues be (if we want to distinguish mass and wave function renormalization explicitly, we have to label edges accordingly
for these external structures \cite{ckI} without essentially changes to our set-up)
\beqa
R=\{\, \includegraphics[scale=0.07]{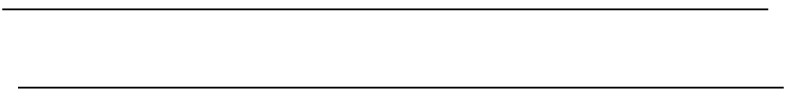}, \includegraphics[scale=0.07]{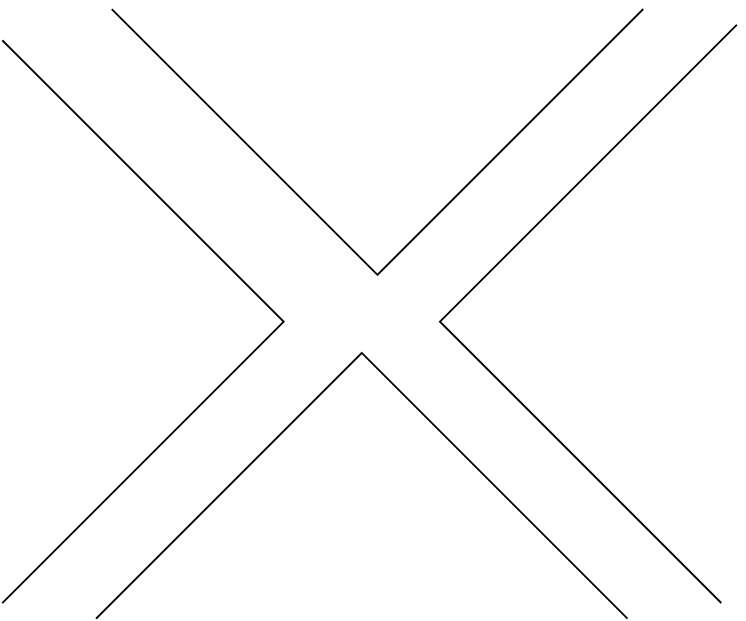}\}.
\eeqa
The insertion operation is defined as the bilinear map
\beqa
\label{insertie}
\Gamma_1 \circ \Gamma_2 := \sum_{\Gamma} n (\Gamma_1, \Gamma_2, \Gamma) \Gamma.
\eeqa
The coefficient $n(\Gamma_1, \Gamma_2, \Gamma)$ counts the number of ways to shrink $\Gamma_2$ 
to its residue in the graph $\Gamma$ such that $\Gamma_1$ is obtained. 

Let us first deal with the insertions of planar regular graphs. We consider here insertions of a $4-$point function, the $2-$point 
function insertions being easier and thus left to the reader. 

\begin{figure}
\centerline{\epsfig{figure=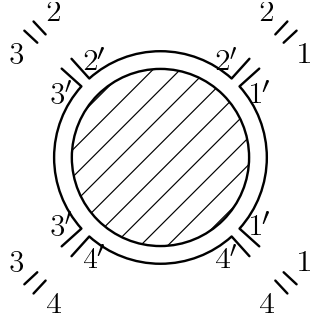,width=4cm} }
\caption{A $4-$point planar regular graph to be inserted in some Moyal vertex. Since all the external edges are on the same face (the external) one, we can define a cyclic ordering.}\label{4p}
\end{figure}

\begin{figure}
\centerline{\epsfig{figure=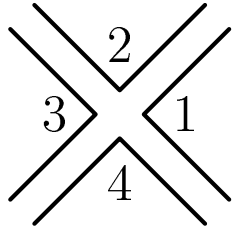,width=3cm} }
\caption{The Moyal vertex where the insertion is made has a cyclic ordering symmetry of its edges. One can thus realize the bijection with the external edges of the graphs to be inserted.}\label{vertex}
\end{figure}

\begin{figure}
\centerline{\epsfig{figure=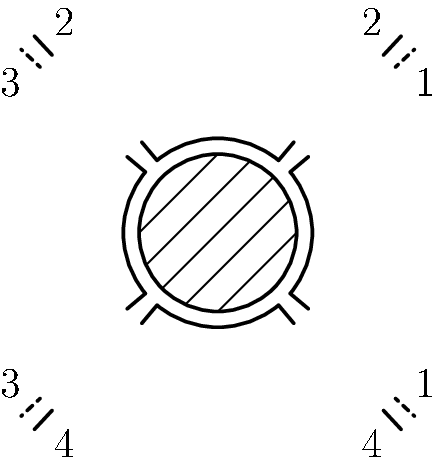,width=4cm} }
\caption{Insertion of a $4-$point planar regular graph with gluing data respecting the cyclic ordering}\label{respect}
\end{figure}

\begin{figure}
\centerline{\epsfig{figure=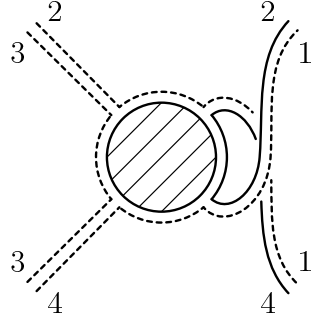,width=4cm} }
\caption{Insertion of a $4-$point planar regular graph with gluing data not respecting the cyclic ordering}\label{disrespect}
\end{figure}

Since the graph to be inserted is considered regular, 
one has all the external edges breaking the same (external) face. We can thus define the before-mentioned  
cyclic ordering on these external edges (see Fig. \ref{4p}).
One can then establish a bijection between these external edges and the edges of the Moyal vertex where the insertion will be made 
(the {\it gluing data}, see Fig. \ref{vertex}).
This gluing data can either
\begin{enumerate}
\item respect the cyclic ordering (see Fig. \ref{respect}) or
\item do not respect the cyclic ordering (see Fig. \ref{disrespect})
\end{enumerate}

We now denote by

\beqa
\label{insertiec}
\Gamma_1 \circ_c \Gamma_2 := \sum_{\Gamma} n (\Gamma_1, \Gamma_2, \Gamma) \Gamma.
\eeqa
the insertion as defined in \eqref{insertie}.
The sum on the right is always over regular graphs, and hence only insertions which respect the cycling ordering contribute.

\begin{figure}
\centerline{\epsfig{figure=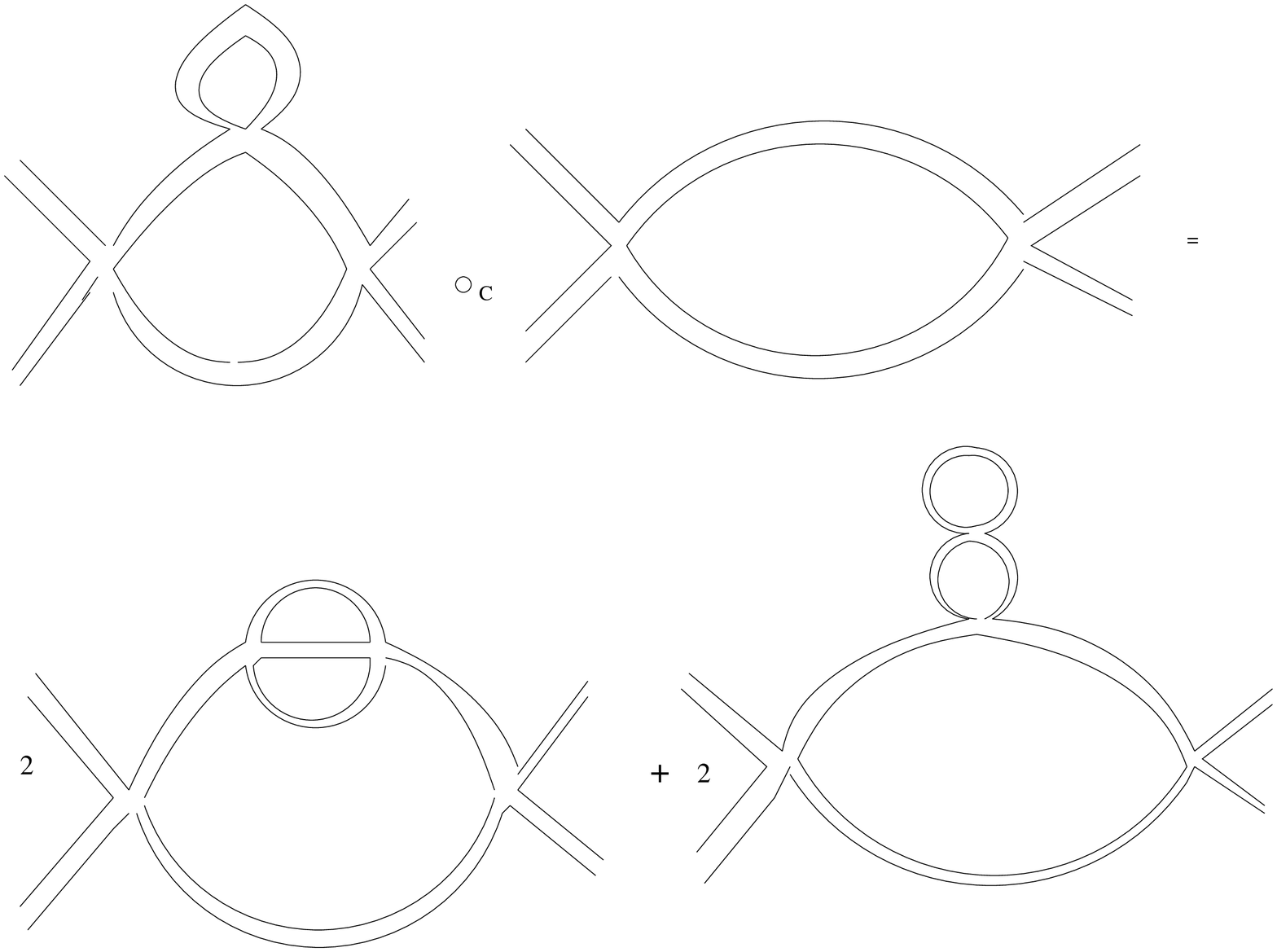,width=8cm} }
\caption{Example of a ribbon graph insertion which respects the cyclic ordering.}\label{ex-insertie}
\end{figure}

\begin{figure}
\centerline{\epsfig{figure=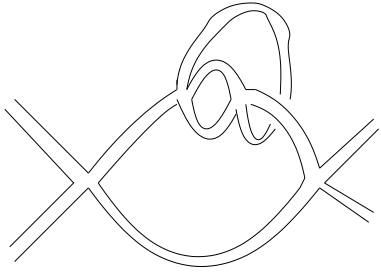,width=5cm} }
\caption{Result of the insertion of Fig. \ref{ex-insertie} if one uses some gluing data not respecting the cyclic ordering.}\label{ex-disrespect}
\end{figure}

Let us illustrate all this by working out the explicit example of Fig. \ref{ex-insertie}.
One can also consider some gluing data not respecting this cyclic ordering (as in Fig. \ref{disrespect}). 
The result is shown in Fig. \ref{ex-disrespect}. 
Note that in the case of commutative $\phi^4$ theory the first graph in the RHS of Fig. \ref{ex-insertie} 
and the graph of Fig. \ref{ex-disrespect} are equivalent. 
Indeed, one can use the symmetry under permutation of the incoming/outgoing fields of the local vertex to 
rewind the non-planar graph of Fig. \ref{ex-disrespect} to the planar regular 
one of Fig. \ref{ex-insertie}. This is not allowed in NCQFT because of the restricted symmetry of the Moyal vertex.

\begin{figure}
\centerline{\epsfig{figure=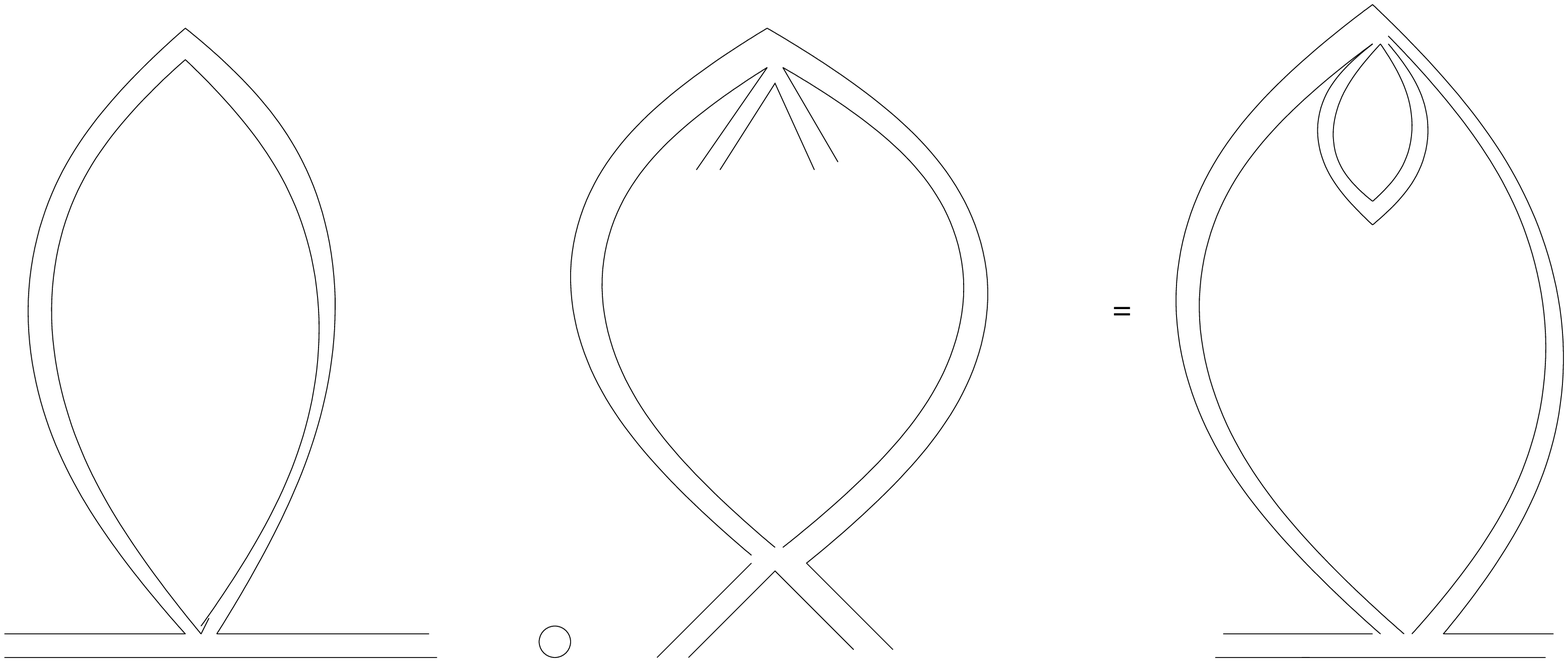,width=9cm} }
\caption{An example of an insertion of a planar irregular graph leading to a planar regular graph. The number of broken faces is thus reduced.
We have inserted a planar irregular (thus convergent) graph in some planar regular one. The result is again planar regular.}\label{reducere}
\end{figure}

Furthermore, in NCQFT another type of insertion is possible: one can insert a $4-$point graph which is planar irregular. 
Not having all its external edges on the same face, one cannot define anymore some cyclic ordering on them. 
One can insert such a graph in a Moyal vertex in a way that reduces the number of broken faces and does not increase the genus of the resulting graph.
This becomes clear in the example of Fig. \ref{reducere}.
Inserting planar irregular graphs into planar graphs in a way that a planar (regular or irregular) graph is obtained is also possible for $4-$point graphs with $3$ broken faces, in a similar way. This is still possible because one has two legs on a particular face.

\begin{figure}
\centerline{\epsfig{figure=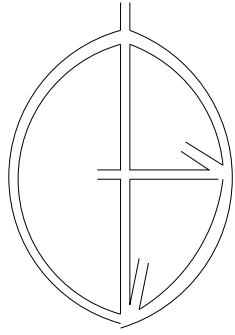,width=3cm} }
\caption{An example of $4-$point graph with four faces broken by external legs. Each face is broken by a single line. For this reason, when inserting such a graph in some Moyal vertex, the resulting graph is non-planar.}
\label{B4}
\end{figure}

Nevertheless this is no longer possible for $4-$point planar irregular graphs with $4$ faces broken by external legs (consider for example the graph of Fig. \ref{B4}). 
This comes from the fact that one does not have anymore $2$ legs on the same face, whose presence, by a proper defined gluing data, could have prevented the final graph to become non-planar. 
One can thus conclude that in any way such a graph is inserted into a Moyal vertex the genus of the resulting graph increases.

As already mentioned in the previous subsection, the same situation occurs when one inserts a planar irregular $2-$point graph (like for example the non-planar tadpole graph of Fig. \ref{nptad}) into some planar graphs.

We will come back on this important issue in the sequel.
Note that such phenomenae are irrelevant in commutative theories, as explained above.

We close this section by a short remark on the accompanying Lie algebras. 
Considering the gluing data compatible with the cyclic ordering, one has an obvious  pre-Lie algebra structure.
Antisymmetrizing the pre-Lie product gives a Lie bracket
\beqa
\label{lie}
[\Gamma_1, \Gamma_2]=\Gamma_1 \circ_c \Gamma_2 - \Gamma_2 \circ_c \Gamma_1,
\eeqa
which defines a Lie algebra structure $L$.
Consider now the graded dual of the universal enveloping algebra of this Lie structure. This gives the renormalization Hopf algebra, 
an algebra which is described in the subsection \ref{sect:hopf}. 
The fact that the products \eqref{insertie} and resp. \eqref{lie} are indeed pre-Lie and resp. Lie products can thus be 
seen as a direct consequence of the existence of the Hopf algebra structure described in the following subsection.

\subsection{Definition of ribbon graph Hopf algebras}
\label{sect:hopf}

Let now the unital associative algebra freely generated by $1$PI non-commutative Feynman graphs 
(including the empty set, which we denote by $1_\ch$). 
The product $m$ is bilinear, commutative and given by the operation of disjoint union. 

We first define a core Hopf algebra 
${\cal H}_{1PI}$, 
which is a straightforward generalization of the core Hopf algebra defined for a commutative theory in \cite{bkultim}. See \cite{core,WvSDK}
for further details.
The coproduct is  defined as
\beqa
&& \Delta:\ch\to\ch\otimes\ch, \nonumber\\
&& \Delta \Gamma = \Gamma \otimes1_\ch + 1_\ch\otimes \Gamma + \sum_{\gamma\subset\Gamma \; 1PI} \gamma \otimes \Gamma/\gamma, \ \forall\Gamma\in\ch_{1PI}.
\label{coproductc}
\eeqa
The definitions of the counit and antipode follow directly and one can easily check that $\ch_{1PI}$ is a Hopf algebra
with all its cohomological richness (see for example \cite{bkultim,WvSDK}).

\medskip

\noi
We then define the renormalization Hopf algebra $\ch$ as follows. Let the coproduct
\beqa
\label{Delta}
&& \Delta:\ch\to\ch\otimes\ch, \nonumber\\
&& \Delta \Gamma = \Gamma \otimes1_\ch + 1_\ch\otimes \Gamma + \sum_{\gamma\subset\underline\Gamma \; \mathrm{planar,\; regular\; or\; irregular}}
 \gamma \otimes \Gamma/\gamma, \ \forall\Gamma\in\ch,\\
&& \Delta \Gamma = \Gamma \otimes1_\ch + 1_\ch\otimes \Gamma + \sum_{\gamma \; \mathrm{planar, \; regular\; or\; irregular}}n(\gamma,\Gamma/\gamma,\Gamma) \gamma \otimes \Gamma/\gamma, \ \forall\Gamma\in\ch.
\label{coproduct}
\eeqa
Here, in the first form, the sum runs over proper subsets of $\Gamma$ ($2-$ or $4-$point graphs) which form disjoint unions of 1PI planar graphs, and over all distinct 
disjoint unions of 1PI planar graphs ($2-$ or $4-$point ones) in the second form. The section coefficient $n(\gamma,\Gamma/\gamma,\Gamma)$ coincides with the one we had before.

This coproduct allows for irregular subgraphs on the left. It is hence able to take account of the finite renormalizations which come with such subgraphs.
For the renormalization of the proper divergent graphs in our theory it suffices to divide by an (co)-ideal which eliminates irregular graphs as described below.

In the commutative case, the core Hopf algebra $\ch_{1PI}$ contains  the renormalization Hopf algebra $\ch$ as a 
quotient algebra (see \cite{bkultim, core}). That holds similarly here, before and after division by that (co)-ideal.
 
Furthermore, let us remark that the renormalization coproduct \eqref{coproduct} is conceptually different from 
the core coproduct \eqref{coproductc}, both in the commutative or the noncommutative case. One can find (ribbon) graphs 
which have non-trivial coproducts in the core Hopf algebra and are in the same time primitive elements in the renormalization Hopf algebra. Indeed, the core Hopf algebra stores more information then the renormalization one 
(for a more detailed analysis of this aspect in the commutative setting, the interested reader may report himself again to \cite{bkultim, core}).

Let now the counit be
\beqa
&&\varepsilon:\ch\to\KK,\nonumber\\
&&\varepsilon (1_\ch) =1,\ \varepsilon (\Gamma)=0,\ \forall \Gamma\ne1_\ch.
\eeqa
Finally the antipode is given recursively by
\beqa
&&  S:\ch\to\ch\label{eq:Antipode}\\
&& S(1_\ch)=1_\ch, \ \ \  S(\Gamma)=-\Gamma-\sum_{\gamma}S(\gamma)\Gamma/\gamma,\nonumber
\eeqa
with the sum taken from the definition of the coproduct.

Let us emphasize that the factorization phenomena appearing in the definition (\ref{coproduct}) of the coproduct $\Delta$ 
corresponds to the renormalizability proved in \cite{noi}.


We can thus state the main result of this subsection:

\begin{theorem}
The pair $({\cal H}, \Delta)$ is a Hopf algebra.
\end{theorem}

We call ker$\, \epsilon$ the {\bf augmentation ideal}. Note that the quantum world, {\it i.e.} all graphs containing loops,  belong to the augmentation ideal.

Let the projection
\beqa
P:\ch\to {\rm ker} \, \e, \ \ \ 
P=id - 1_\ch \e
\eeqa
and
\beqa
{\rm Aug}^{\otimes k} = (P\otimes P \ldots \otimes P)\Delta^{k-1}.
\eeqa
We define 
$$|\Gamma|_{\rm aug} $$
to be the {\bf augmentation degree}.

Let us also denote by
$$|\Gamma| $$
the number of independent loops of some graph $\Gamma$.

The $1$PI graphs $\Gamma$  provide the linear generators $\delta_\Gamma$. 
The Hopf algebra is an algebra, the free commutative (but not co-commutative) algebra of these generators.
We denote 
\beqa
\ch_{\rm lin}={\rm span}\, (\delta_\Gamma).
\eeqa

Note that, as in the commutative case one can also define a Hopf algebra of decorated rooted trees $\ch_{\rm rt}$ (see for example \cite{bk} or \cite{dk}). Furthermore, as explained in \cite{hopf} the formal definition of the Bogoliubov subtraction operator remains the same as in the commutative field theoretical setting.

\subsection{Planar irregular graphs; semi-direct structure}
\label{discard}

In this subsection we come back on the issue of the $2-$point planar irregular graphs, this time from the point of view of the Hopf algebra defined in the previous subsection. 

\begin{proposition}
\label{prop-ideal}
The ideal  ${\cal H}^{\rm{pli}}$ generated by the 1PI $2-$point planar irregular Feynman graphs is a Hopf ideal and coideal in ${\cal H}$,
\beqa
\label{ideal}
\Delta ({\cal H}^{\rm{pli}})\subseteq {\cal H}^{\rm{pli}} \otimes {\cal H} +{\cal H} \otimes {\cal H}^{\rm{pli}}, \ \varepsilon ({\cal H}^{\rm{pli}})=0, S({\cal H}^{\rm{pli}})\subseteq {\cal H}^{\rm{pli}}.
\eeqa
\end{proposition}
{\it Proof.} 
Let us consider the non-trivial part of the coproduct $\Delta$. We denote the interior broken face by $f$.
If one chooses some $2-$ or $4-$point subgraph of $\Gamma$ to contain the face $f$, then the respective subgraph will be planar irregular and hence not primitive divergent. Thus, all the primitive divergent subgraphs must not contain the respective face $f$. This face will therefore be retrieved in the cograph, which leads to
\beqa
\Delta' ({\cal H}^{\rm{pli}})\subseteq {\cal H} \otimes {\cal H}^{\rm{pli}},
\eeqa
Taking also the trivial part of the coproduct one obtains \eqref{ideal}. Note that when applying the coproduct on these ribbon graphs, 
the number of internal faces conserves as a sum of the number of internal faces of the graph and of the cograph. (QED)

\medskip

A direct consequence of this Proposition is that one can discard 
this planar irregular sector (as we did in the previous subsection) by simply taking the respective quotient  
by the (co)-ideal ${\cal H}/{\cal H}^{\rm{pli}} $. This cannot however be 
done for the non-planar sector as well, because the non-planar sector does not form a Hopf coideal (one can easily find some counterexamples).
We can eliminate it though by working in a suitable quotient Hopf algebra.

Furthermore, let us emphasize on the fact that on the Proposition \ref{prop-ideal} above we have dealt with both the notion of 
Hopf {\it co}ideal (implying the use of the coproduct $\Delta$) and the (here trivial) notion of ideal (implying the use of the product $m$).

\medskip

We have thus established in this section three distinct Hopf algebra structures:
\begin{enumerate}
\item the core Hopf algebra $\ch_{1PI}$ given by \eqref{coproductc},
\item the Hopf algebra $\ch$ given by \eqref{coproduct},
\item the Hopf algebra obtained from $\ch$ by diving with the (co)-ideal eliminating the planar irregular graphs (as described above).  
\end{enumerate}

Let us end this section by giving the semi-direct structures of the Lie algebras associated to the three cases above:

\begin{proposition}
\label{sd}
 The Lie algebra $L$ is the semi-direct product of the abelian Lie algebra $L_0$ by $L_c$, where with respect to the three cases enumerated above one has:
\begin{enumerate}
\item $L_c$ is $\ch_{1PI}$ and $L_0$ is the empty set;
\item $L_c$ is generated by the  planar regular $2-$ and $4-$point graphs, planar irregular $4-$point graphs with two or three broken faces
and $L_0$ is   generated by the $6,8,\ldots-$point graphs (planar regular and irregular) as well as the $2-$point planar irregular graphs and $4-$point planar irregular graphs with four broken faces;
\item $L_c$ is generated by the planar regular $2-$ and $4-$point graphs and $L_0$ is generated by the $6,8,\ldots-$point graphs (planar regular).
\end{enumerate}
\end{proposition}
{\it Proof.} 
In the first case, there is no difference with the commutative case; the result stated above is a straightforward consequence of the fact that vertices of any valence are allowed to appear, as opposed to usual Feynman graphs in renormalizable perturbative quantum field theories (see \cite{WvSDK} for more details).
The second case is the most involved one. One expects to have the $L_c$ part generated simply by the planar regular and irregular $2-$ and $4-$point graphs. Nevertheless, we have seen in subsection \ref{preLie} that the planar irregular $2-$point graphs and the planar irregular $4-$point graphs with four broken faces cannot be inserted without increasing the genus (thus leading to non-planarity). Finally, the last case is treated along the same lines as for the renormalization Hopf algebra of commutative theories (see \cite{ckI}) since we deal here only with {\it planar regular} graphs (and with insertions respecting the cyclic ordering, as explained in subsection \ref{preLie}). (QED)

\section{More on ribbon graphs}
\renewcommand{\theequation}{\thesection.\arabic{equation}}   
\setcounter{equation}{0}
\label{more}

\subsection{Symmetry factor of graphs}

\begin{figure}
\centerline{\epsfig{figure=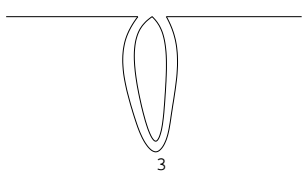,width=4cm} }
\caption{The down tadpole. This graph is planar regular and hence primitive. Its symmetry factor is equal to $1$.}
\label{down-tadpole}
\end{figure}

\begin{figure}
\centerline{\epsfig{figure=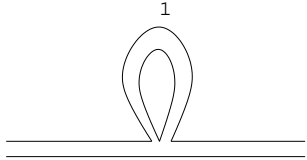,width=4cm} }
\caption{The up tadpole. This graph is planar regular and hence primitive.}
\label{up-tadpole}
\end{figure}

The symmetry factor of a graph $\Gamma$ is defined as the rank of automorphism group of $\Gamma$.
The use of ribbon graph changes the picture with respect to the $\phi^4$ theory. One has, for example a symmetry factor of $1$ for the down (resp. up) tadpoles of Fig. \ref{down-tadpole} (resp. Fig. \ref{up-tadpole}).

\begin{proposition}
The symmetry factor of a ribbon graph in NCQFT is equal to $1$.
\end{proposition}
{\it Proof.} We proceed by induction on the number of loops $b$. The affirmation is obviously true for $b=1$. To obtain graphs we can insert $2-$ and resp. $4-$point graphs in propagators and resp. Moyal vertices. Thus, if the statement is true for some $b\in \NN$  it will be true for $b+1$ loops. This completes the proof. (QED)

\medskip

Let us remark that this Proposition is a direct consequence of the fact that the Moyal vertex is symmetric only under cyclic permutation of the incoming/outgoing fields (see above). In the case of commutative field theory it is the symmetry under the total group of permutations of the vertex that is responsible for non-unit symmetry factors.

\subsection{Permutation of external edges}

\begin{definition}
Let 
$$|\Gamma|_V$$
be the number of distinct ribbon graphs $\Gamma$ which are equal upon removal of external edges. 
\end{definition}

\begin{figure}
\centerline{\epsfig{figure=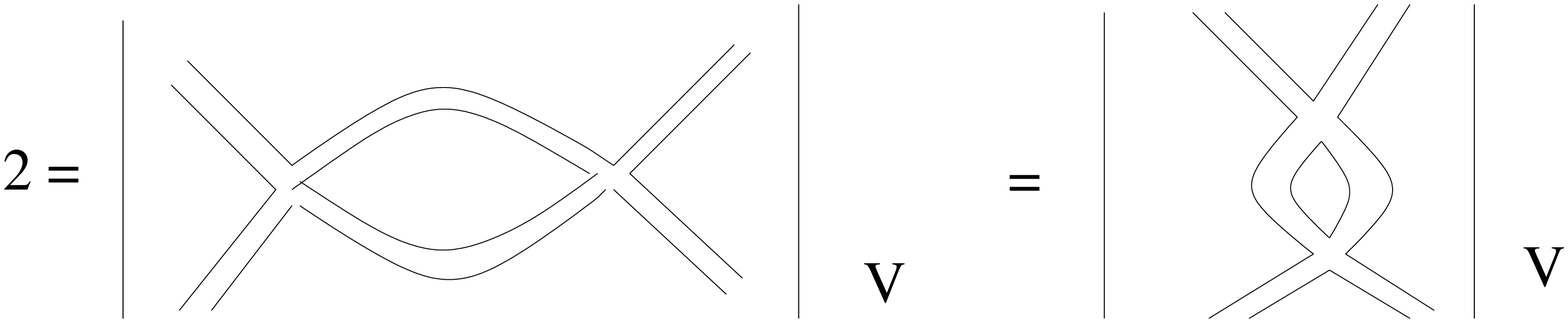,width=9cm} }
\caption{An example for $\Gamma_V$.}
\label{permutation}
\end{figure}

Let us consider the example of Fig. \ref{permutation}.
Note that in the case of commutative $\phi^4$ theory one has $|\Gamma|_V=3$ for the example above (see \cite{anatomy}). The missing third graph disappears because we consider only the planar regular sector of the theory.

\subsection{Number of maximal forests}

\begin{definition}
The number of maximal forests ${\rm maxf}$ of a graph $\Gamma$ is the number of ways to shrink subdivergencies to Moyal points such that the resulting cograph is primitive.
\end{definition}

\begin{figure}
\centerline{\epsfig{figure=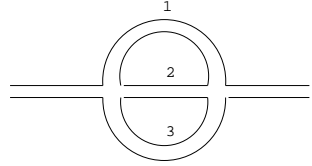,width=4cm} }
\caption{The sunshine graph, with three internal lines. Its maximal number of forests is equal to $2$.}
\label{sunshine}
\end{figure}

\begin{figure}
\centerline{\epsfig{figure=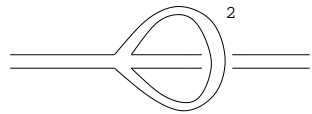,width=4cm} }
\caption{The ``non-planar'' tadpole obtained from shrinking the bubble graph formed by lines $1$ and $3$ in the sunshine graph of Fig. \ref{sunshine}}
\label{np-tadpole}
\end{figure}

Let us now give an example which illustrates the difference in calculating ${\rm maxmf}$ with respect to the commutative $\phi^4$ graphs.
The graph of Fig. \ref{sunshine}
has a maximal number of forests equal to two. Indeed, when the subdivergence is represented by the bubble graph formed of the internal lines $1$ and $2$ then the resulting cograph is the down tadpole of Fig. \ref{down-tadpole}.
If the subdivergence is taken to be the bubble graph formed of the internal lines $2$ and $3$ then the resulting cograph is up tadpole  of Fig. \ref{up-tadpole}.
Finally, if one takes the divergence to be given by the internal lines $1$ and $3$ then the resulting cograph is the ``non-planar'' tadpole  of Fig. \ref{np-tadpole}.
which is not primitive. Hence, $\rm maxf$ is equal to $2$. In the case of the commutative $\phi^4$ theory, the three tadpoles above are equivalent 
and thus the number of maximal forests will be equal to $3$.

\subsection{Number of bijections when gluing graphs; number of insertion places}

We denote by {\bf bij}$(\gamma_1, \gamma_2, \gamma)$ the number of bijections between the external edges of $\gamma_2$ and adjacent edges of places $p~${\bf res}$(\gamma_2)$ in $\gamma_1$ such that $\gamma$ is obtained.

We call (subsets of) edges and vertices of a graph {\bf places} of the respective graph. 
We also use the notation $[\gamma | X]$ for the number of insertion places of the graph $X$ in $\gamma$.

The values taken by this parameter do not change if one deals with ribbon graphs. Let us end this section by recalling that in \cite{anatomy} other parameters on Feynman graphs have been defined. All these generalize to ribbon graphs (one just needs to take care of the differences like the one we saw already here). We do not need in this paper these other notions so we do not deal with them here.

\section{Hochschild cohomology in NCQFT: Moyality and Dyson-Schwinger equations}
\renewcommand{\theequation}{\thesection.\arabic{equation}}   
\setcounter{equation}{0}

We  use in this section the Hochschild cohomology of the previous Hopf algebras 
to prove that one can absorb the singularities of such a NCQFT in ``Moyal''-like counterterms ({\it i. e.} counterterms of the same form of the ones in the original action, see above). Indeed, every divergent graph $\gamma$ without subdivergencies determines a Hochschild $1-$cocycle $B_+^\gamma$ (see below). Furthermore, any relevant graph is in the range of such a $1-$cocycle. This ensures that any relevant term in the perturbative expansion is in the image of a Hochschild $1-$cocyle; this allows to prove ``Moyality'' (see for example \cite{bk}, where ``locality'' is just replaced by ``Moyality'', because the formal definition of the Bogoliubov operator remains the same).

We also study the Dyson-Schwinger equation as formal construction based on the Hochschild cohomology of these Hopf algebras. Thus one can state that the Hochschild cohomology leads the way from perturbative to non-perturbative physics. 
This construction extends the one of commutative field theories (see for example \cite{bk, anatomy}). Throughout this section we generally follow closely the results in \cite{bk, dk, anatomy}.

\medskip

Before going further, let us state here that these results hold for the algebraic structures associated to both types of renormalizable NCQFT models introduced in section \ref{rappel}.

\medskip

 Let us firstly recall (following \cite{dk}) some useful definitions regarding the Hochschild cohomology.
Let $A$ 
be a bialgebra and $\Delta$ its coproduct. We regard linear maps
$$ L: A \to A^{\otimes n} $$
as $n-$cochains. We define a coboundary map $b$,
$$ b^2=0,$$
by
\beqa
bL := (id \otimes L) \circ \Delta + \sum_{i=1}^n (-1)^i \Delta_i \circ L +
(-1)^{n+1} L \otimes \One,
\eeqa
where $\Delta_i$ denotes the coproduct applied to the $i-$th factor in $A^{\otimes n}$. This defines the cohomology of $A$.

\bigskip

Let now $(B_+^{d_n})_{n\in\NN}$ a set of Hochschild $1-$cocycles on such a Hopf algebra.
The Dyson-Schwinger equation writes
\beqa
\label{ds}
X=\One + \sum_{n=1}^\infty \omega_n \lambda^n  B_+^{d_n} (X^{n+1})
\eeqa
in ${\cal H}[[\lambda]]$. The parameters $\omega_n$ are scalars. 
One decomposes the solution as
\beqa
\label{ansatz}
X=\sum_{n=0}^\infty \lambda^n c_n, {\mbox{ with }} c_n\in \cal H.
\eeqa
One has to sum up on the contribution corresponding to the planar regular as well to the planar irregular ribbon graphs, since the latter sector can lead to (planar regular) contributions when the operator $B_+$ interferes  (see subsection \ref{preLie}). This is a major difference with respect to the commutative case, because in a commutative framework this distinction is irrelevant and one does not has to deal with this type of phenomenas. We can illustrate this by explicitly splitting
\beqa
\label{spliting}
c_n=c_n^{{\rm reg}}+\widetilde c_n,
\eeqa
where by $c_n^{{\rm reg}}$ we refer to the regular sector and by $\widetilde c_n$ we refer to the irregular sector.

\begin{lemma}
\label{lema} (Lemma $2$ of \cite{bk})
The Dyson-Schwinger equation \eqref{ds} has a solution given by
$c_o=\One$ and 
\beqa
c_{n} = \sum_{m=1}^{n} \omega_m B_+^{d_{m}}\left(\sum
_{k_1+\ldots+k_{m+1}=n-m,\, k_i\ge 0} c_{k_1}\ldots
c_{k_{m+1}}\right).
\eeqa
\end{lemma}
{\it Proof.} As in the commutative case (see \cite{bk}), one needs to insert the ansatz \eqref{ansatz} in the Dyson-Schwinger equation \eqref{ds}. Sorting then by powers in the coupling constant $\lambda$ yield the result. Furthermore, uniqueness is obvious. The use of ribbon graphs does not change the validity of these arguments. 
The only thing that is changed is the fact that one has to sum up also on the planar irregular sector, which leads via the operator $B_+$ to planar regular contributions.
(QED)

\medskip

Let us now switch for the moment to a {\it description in terms of decorated trees}. 
 We denote by $dec(v)$ the decoration and by 
 fert$(v)$ the fertility ({\it i. e.} the number of outgoing edges) of the vertex $v$. Furthermore, the decoration weight of such a tree is defined as the sum of the decorations of the vertices.
We then define the coefficients
\begin{equation*}
\gamma_v= \left\{ \begin{array}{rl}
\omega_{|dec(v)|}\frac{(|dec(v)|+1)!}{(|dec(v)|+1-\operatorname{fert}(v))!}
&\mbox{ if }\operatorname{fert}(v)\le |dec(v)|+1 \\ 0 & \mbox{
else}.
\end{array} \right.
\end{equation*}
For such a decorated tree, the coefficient
\beqa
\label{coef1}
\prod_v \gamma_v
\eeqa
can be interpreted by considering every decorated tree as an operadic object with  $|dec(v)|+1-\operatorname{fert}(v)$ inputs at each vertex $v$. The total number of inputs is $n+1$, where $n$ is the decoration weight of the respective tree. The coefficient \eqref{coef1} is the number of planar ({\it i.e.} noncommutative) embeddings of this operadic tree when keeping the trunk ({\it i.e.} the original tree) fixed.
Note that the sense of ``planar'' and resp. ``noncommutative'' here refers to the  decorated trees and not to the Feynman graphs or resp. spacetime (as used in the rest of the paper).
For an explicit example of the interpretation above, one can refer himself to \cite{bk}.

This type of reasoning is not dependent on the fact that the respective decorated tree was obtained from a ribbon or an ``usual'' commutative Feynman graph.

Before going further let us remark that the operation of inserting graphs into graphs can be mathematically written down in an {\it operadic language}.

\begin{theorem}
\label{teorema} (Theorem $3$ of \cite{bk})
The elements $c_n\in{\cal H}$ generate a Hopf subalgebra in ${\cal H}$
\beqa
\Delta (c_n)=\sum_{m=1}^n P_k^n \otimes c_n
\eeqa
where $P_n^k$ are polynomials of degree $n-k$ in the elements $c_\ell$, $\ell\le n$, given by
\beqa
\label{enunt}
P^{n}_k = \sum_{l_1+\ldots+l_{k+1}=n-k}c_{l_1}\ldots c_{l_{k+1}}.
\eeqa
\end{theorem}
{\it Proof.} We give here an operadic proof which follows the one of \cite{bk}. 
Let $\mu_n$ some maps in
$$ {O}^{[n]}: V^{\otimes n} \to V$$
for some space $V$ and $G(\lambda)$ a formal series in $\lambda$ with coefficients in
 ${O}^{[j]}$. We denote the identity map on $V$ as $\One_V$.
As a variation of the Dyson-Schwinger equation \eqref{ds} we write down the operadic fix point equation
\beqa
 G(\lambda)= \One_V + \sum_n \lambda^n \mu_{n+1}(G(\lambda)^{\otimes (n+1)}).
\eeqa
One writes $G(\lambda)=\One_V + \sum_k \alpha^k \nu_k$. By induction it then follows that $\nu_k\in{O}^{[k+1]}$. Furthermore, $G(\lambda)$ is a sum (with unit weights) over all maps which one obtain by composition of some undecomposable maps $\mu_n$.

\medskip

Let us now consider the coproduct of decorated rooted trees and some  monomial 
$$\nu^{r_1}\ldots  \nu_{i_l}^{r_l}.$$
Note that this monomial lives in the PROP $V^{\otimes
(r_1i_1+\ldots+r_li_l+r)}\to V^{\otimes r}$, where $r=\sum_{i=1}^l r_i$. 
The number of ways such a monomial 
can be composed with any element in  $O^{[r-1]}$ is 
\begin{equation}
\label{coef2}
\frac{r!}{r_1!\ldots r_l!}
\end{equation}
This is the contribution to the term in the coproduct which has $\nu_k$ on the RHS and the given monomial on the LHS (because the $\nu_i$ sum over all maps with unit weight). The same argument \eqref{coef2} also determines the coproduct on the $c_k$ on the initial Dyson-Schwinger equation \eqref{ds}:
\begin{equation}
\label{pre}
P^n_k = \sum_{\genfrac{}{}{0pt}{}{i_1r_1+\ldots+i_lr_l=n-k}{0\le
i_s<i_{s+1}\le k,\,\sum r_i=k+1}}\frac{(k+1)!}{r_1!\ldots
r_l!}c_{i_1}^{r_1}\ldots c^{r_l}_{i_{l}}.
\end{equation}
This comes from the fact that the trees in $c_k$ are weighted by the noncommutative (planar) product \eqref{coef1} over the vertices; the coproduct respects this planar structure.

Equation \eqref{pre} is in agreement with \eqref{enunt}.
As above, these arguments apply also in the case of NCQFT, when replacing the Feynman graphs with ribbon Feynman graphs. (QED)

\begin{remark}
The coefficient \eqref{coef1} corresponds to the noncommutative (planar) case while the coefficient \eqref{pre} corresponds to the commutative (non-planar) case. Note that, as in the proof of Theorem \ref{teorema}, the terminology ``commutative'' (resp. ``planar'') is related to the decorated rooted trees and not to spacetime (resp. Feynman graphs).  
\end{remark}

\bigskip

Let us come back at the case of the renormalization Hopf algebra of Feynman graphs. 
As in \cite{anatomy} we define the maps from $\ch$ to $\ch_{{\rm lin}}$
\beqa
\label{b+}
B_+^{k; r}&=&\sum_{{\mid\gamma\mid=k
\atop\mid\gamma\mid_{\rm aug}=1} \atop{\rm
res}(\gamma)=r} B_+^\gamma,\nonumber\\
B_+^\gamma(X)&=&\sum_{\Gamma\in {\cal H}_{\rm lin}}\frac{{\bf bij
}(\gamma,X,\Gamma)}{|X|_\vee}\frac{1}{{\rm
maxf}(\Gamma)}\frac{1}{\left[ \gamma|X\right]}\Gamma,
\eeqa 
for all graphs $X$ in the augmentation ideal. 
Furthermore, we let 
\beqa
\label{trivial}
B_+^\gamma(1_\ch)=\gamma.
\eeqa

This definition ensures that $B_+^{k; r}$ is a Hochschild closed map. This is achieved thanks to the counting of the number of maximal forests. Thus the map $B_+^\gamma$ is a generalization of the pre-Lie insertion into $\gamma$ (see \cite{anatomy} for further details).

Let 
\beqa
\label{def-ckr}
c_k^r= \sum_{|\Gamma|=k\atop {\rm res}(\Gamma)=r} \Gamma
\eeqa
be the sum of graphs of a given loop number and residue and let $M_r$ be the set of graphs such that {\bf res}$(\Gamma)=r$ for all $r\in R$.

Following again \cite{anatomy}, one has the following result:

\begin{theorem} 
\label{toate} (Theorem $5$ of \cite{anatomy})
\beqa 
i)\; \Gamma^{r}\equiv 1+\sum_{\Gamma\in M_r}{\Gamma}=1+\sum_{k=1}^\infty g^k B_+^{k;r}(X_{k,r})
\eeqa 
\beqa
ii)\; \Delta(B_+^{k;r}(X_{k,r}))=
B_+^{k;r}(X_{k,r})\otimes \One
+({\rm id}\otimes B_+^{k;r})\Delta(X_{k,r}).
\eeqa
\beqa 
iii)\; \Delta(c^{r}_{k})=\sum_{j=0}^k {\rm Pol}^{r}_j(c)\otimes
c^{r}_{k-j},
\eeqa 
where ${\rm Pol}^{r}_j(c)$ is a polynomial in the
variables $c_m^r$ of total degree $j$.
\end{theorem}

Let us also mention the important fact that the Hochschild $1-$cocyle $B_+$ above mixes the planar irregular sector with the planar regular sector of the theory by the mechanism showed in subsection \ref{preLie}. Thus, in the result {\it i)} above, this planar irregular sector has to be included in the set of graphs $X_{k,r}$ of loop number $k$ and residue $r$. 
This is a major difference with the case of commutative field theory.

The result $iii)$ ensures that the elements $c_k^r$ form a {\it Hopf subalgebra}.  As it was showed for a commutative field theory in  \cite{exact},
this is of particular importance in the road towards some exact solution of the Dyson-Schwinger equation,

\section{A two-loop example}

To illustrate the theorems of the previous section, we completely work out here a non-trivial two-loop example. 

\subsection{One loop}
\label{1}

For the noncommutative propagator and vertex, one has
\beqa
\label{B+-1l}
B_+^{1,{\includegraphics[scale=0.07]{propa.eps}}}&=&B_+^{{\includegraphics[scale=0.07]{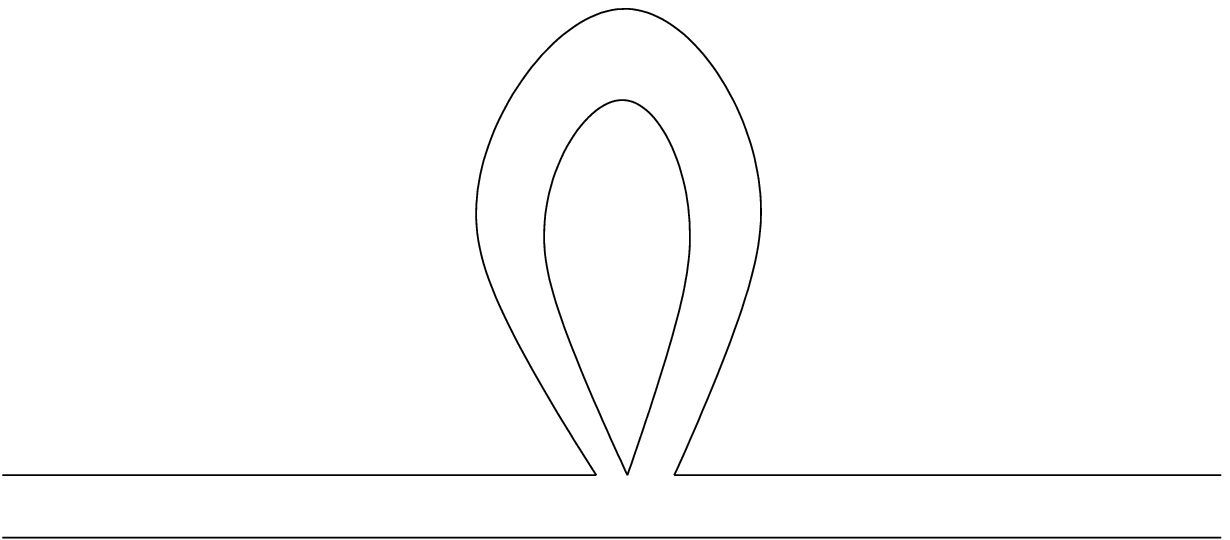}}}+B_+^{{\includegraphics[scale=0.07]{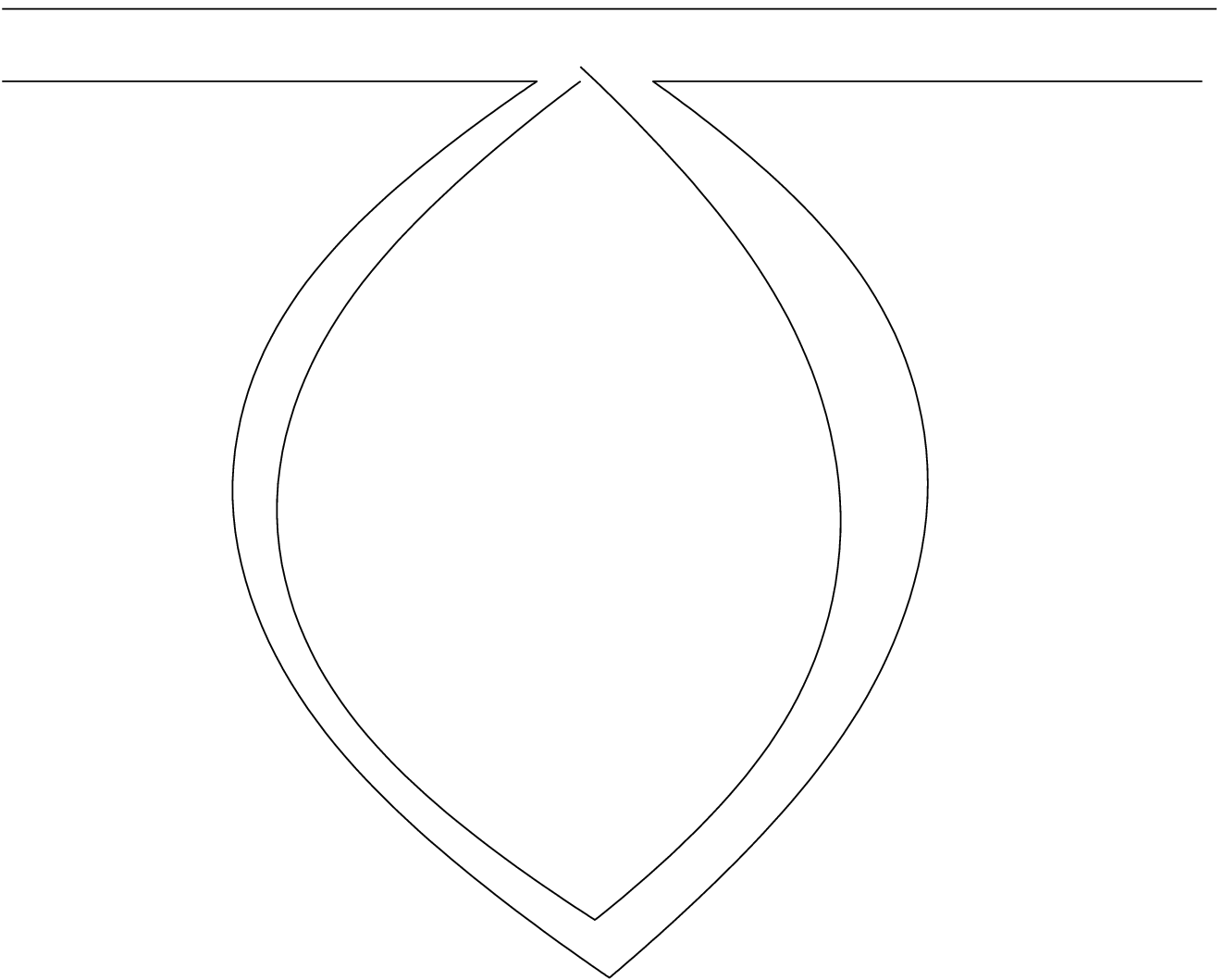}}},\nonumber\\
B_+^{1,{\includegraphics[scale=0.07]{vertex.eps}}}&=&B_+^{{\includegraphics[scale=0.05]{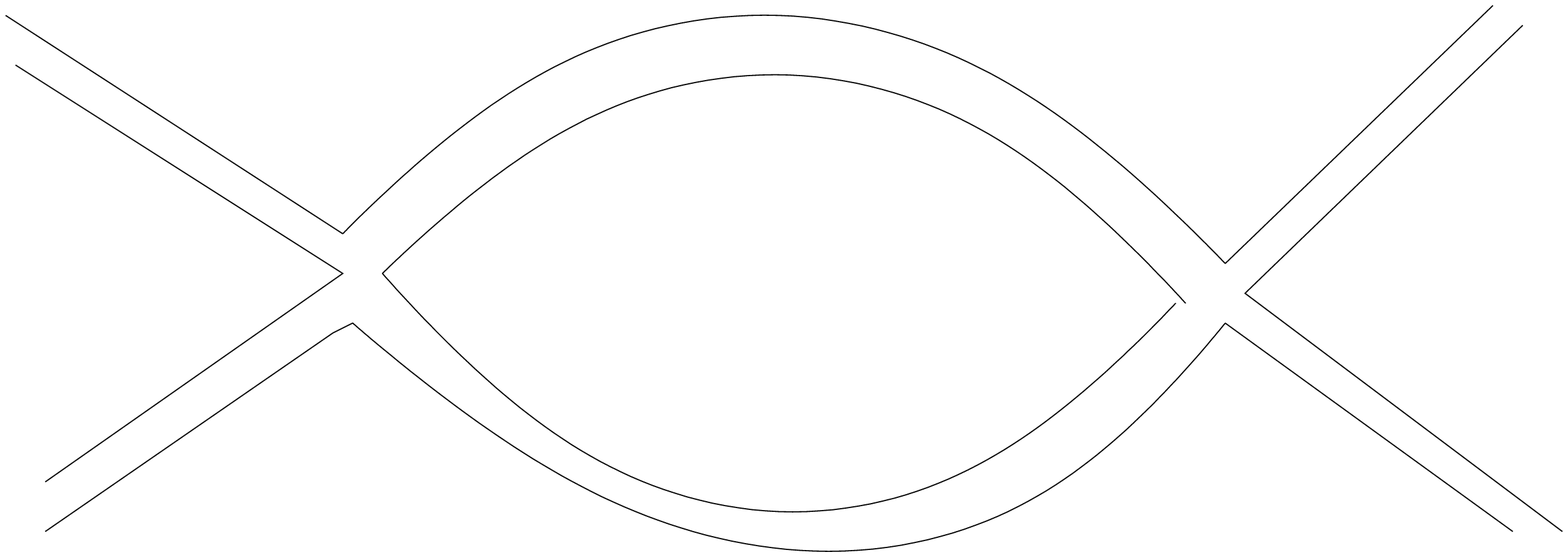}}}+B_+^{{\includegraphics[scale=0.05]{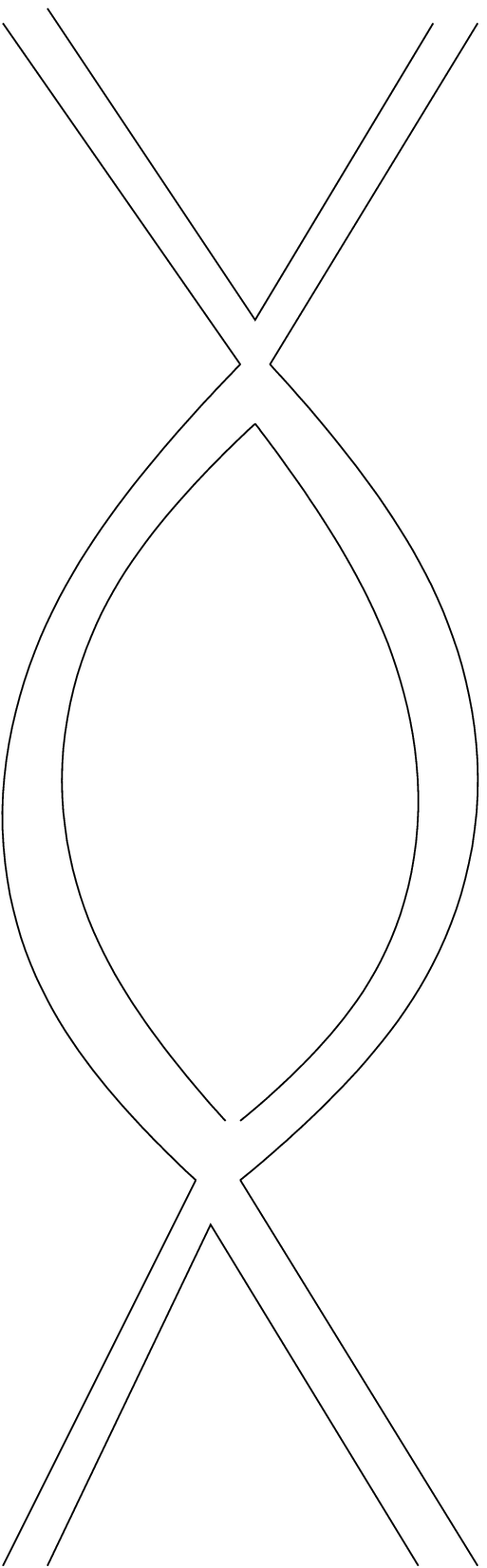}}}.
\eeqa
Applying this map on the unit of the Hopf algebra of graphs $1_\ch$ leads to
\beqa
c_1^{{\includegraphics[scale=0.07]{propa.eps}}}&=&B_+^{1,{\includegraphics[scale=0.07]{propa.eps}}}(1_\ch),\nonumber\\
c_1^{{\includegraphics[scale=0.07]{vertex.eps}}}&=&B_+^{1,{\includegraphics[scale=0.07]{vertex.eps}}}(1_\ch).
\eeqa
Applying \eqref{trivial} and \eqref{B+-1l}  leads trivially to
\beqa
\label{c1uri}
c_1^{{\includegraphics[scale=0.07]{propa.eps}}}&=&{{\includegraphics[scale=0.07]{tadpole-up.eps}}}+{{\includegraphics[scale=0.07]{tadpole-down.eps}}},\nonumber\\
c_1^{{\includegraphics[scale=0.07]{vertex.eps}}}&=&{{\includegraphics[scale=0.05]{vertex-1l-1.eps}}}\ + \ {{\includegraphics[scale=0.05]{vertex-1l-2.eps}}}.
\eeqa

Let us now go further to the more involved case of the two-loop computations.

\subsection{Two loops}
\label{2}

We first work out the easier two-loop two-point function and then proceed with  the four-point one. Using the definition \eqref{def-ckr}, one gets
\beqa
\label{c2}
c_2^{{\includegraphics[scale=0.07]{propa.eps}}}&=&{{\includegraphics[scale=0.1]{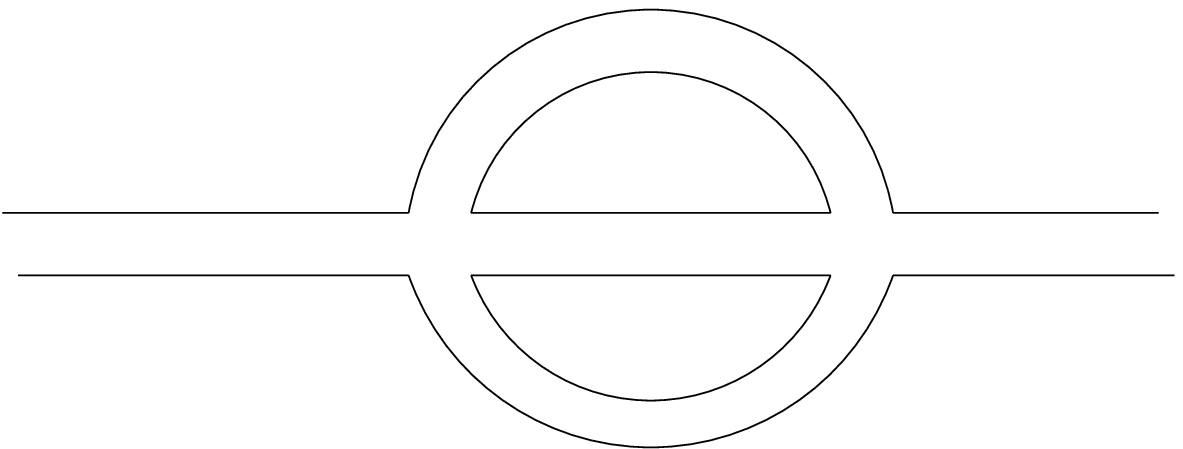}}}\, + {{\includegraphics[scale=0.1]{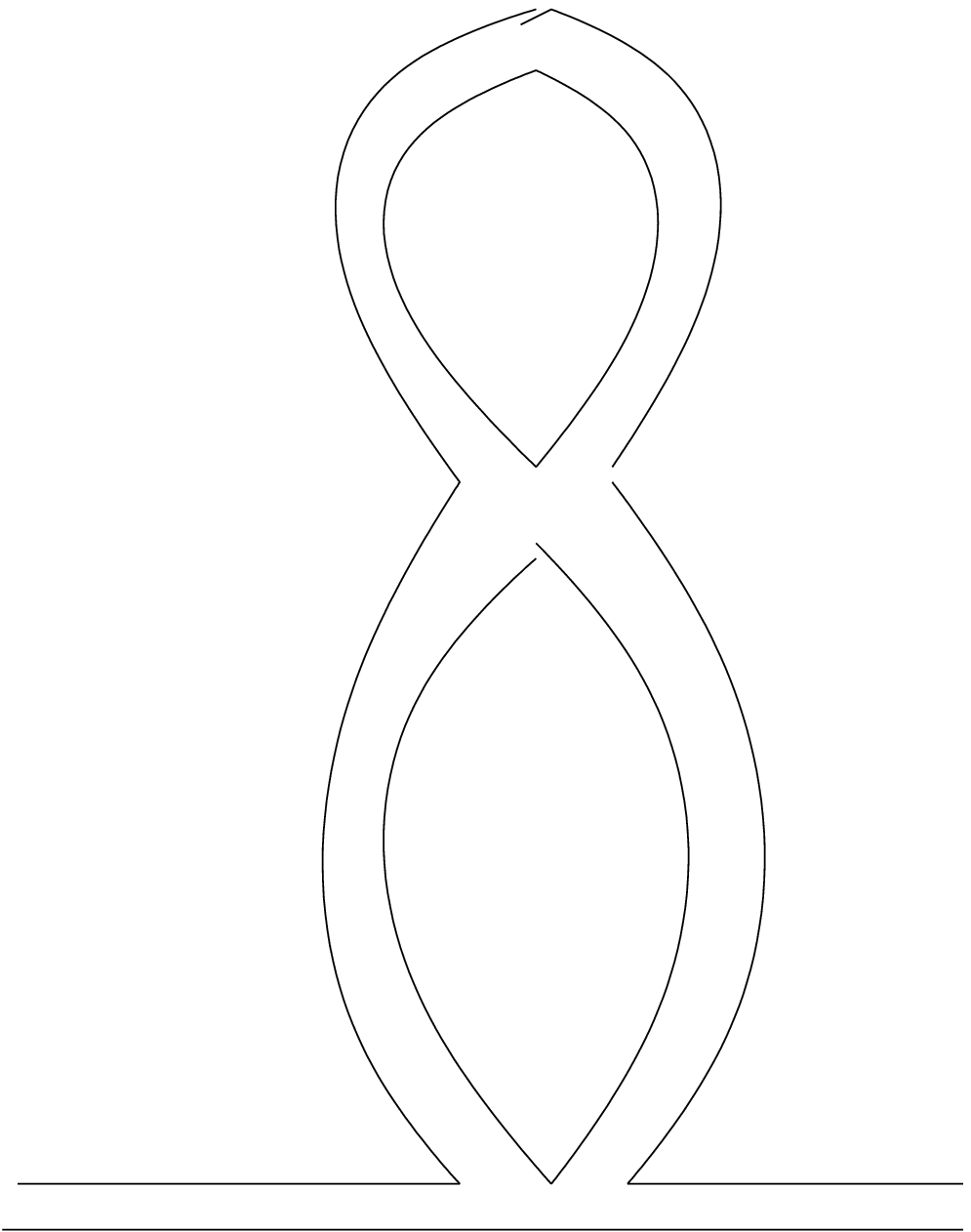}}}\, + {{\includegraphics[scale=0.1]{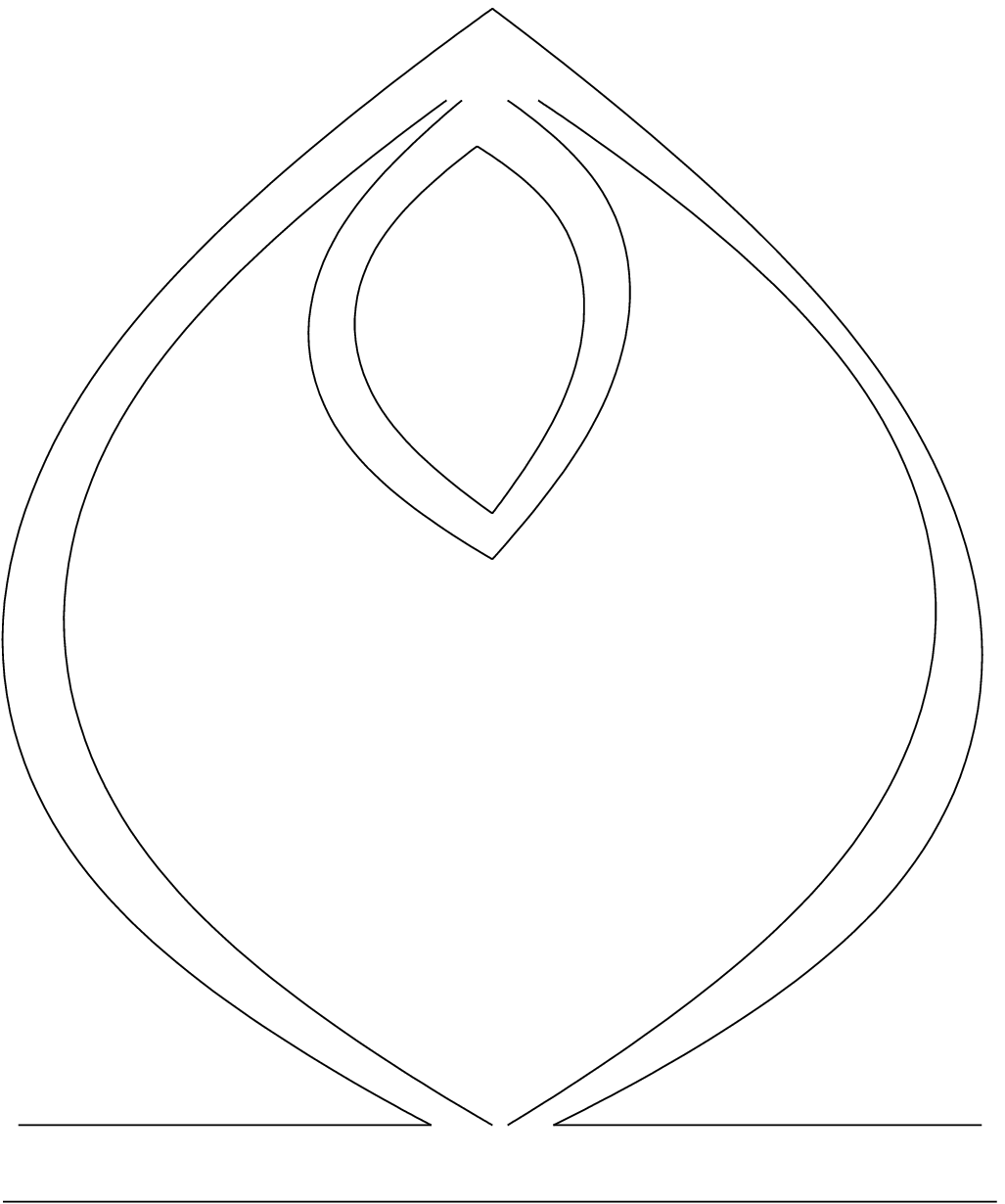}}}\, +  {{\includegraphics[scale=0.1]{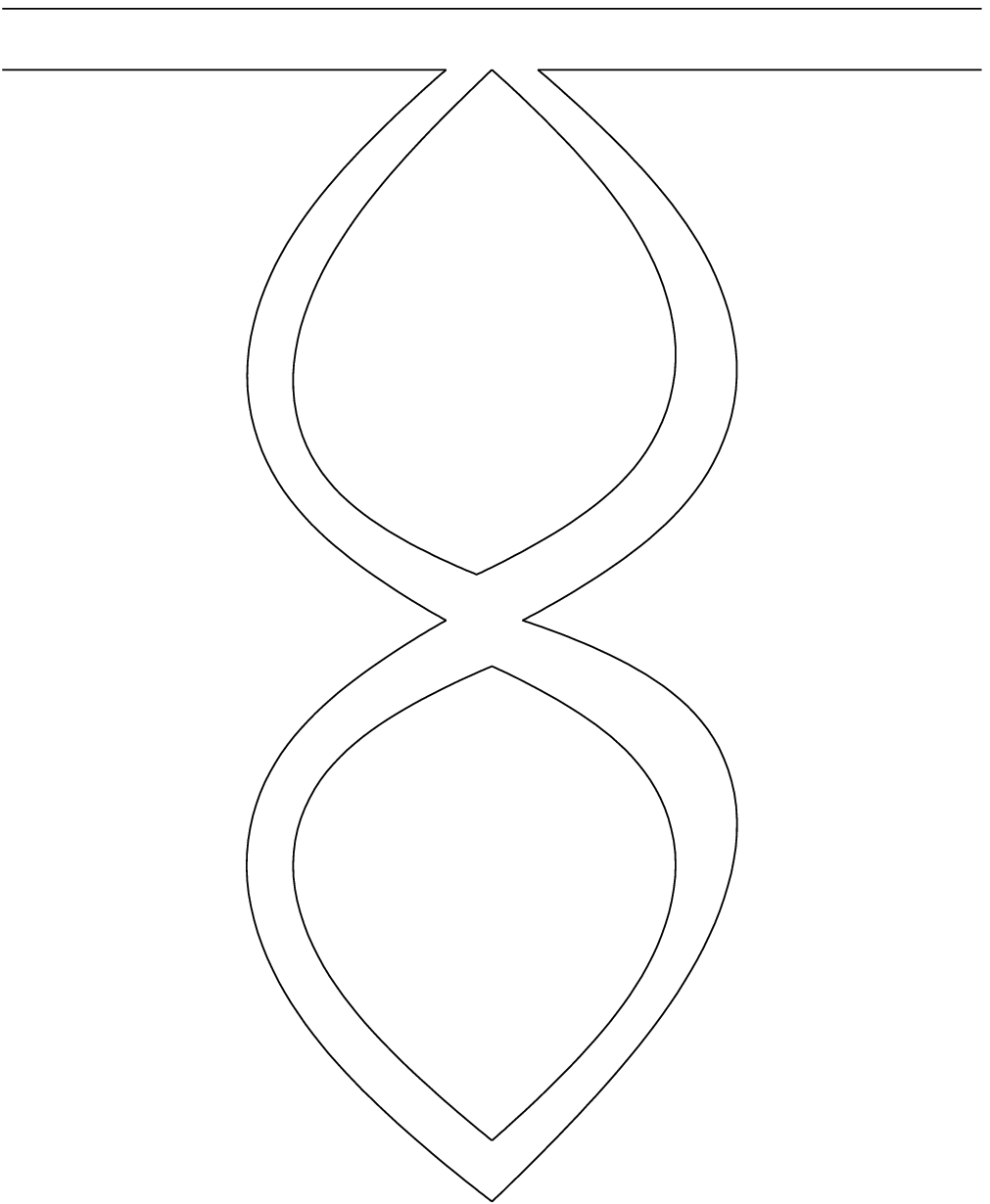}}}\, +  {{\includegraphics[scale=0.1]{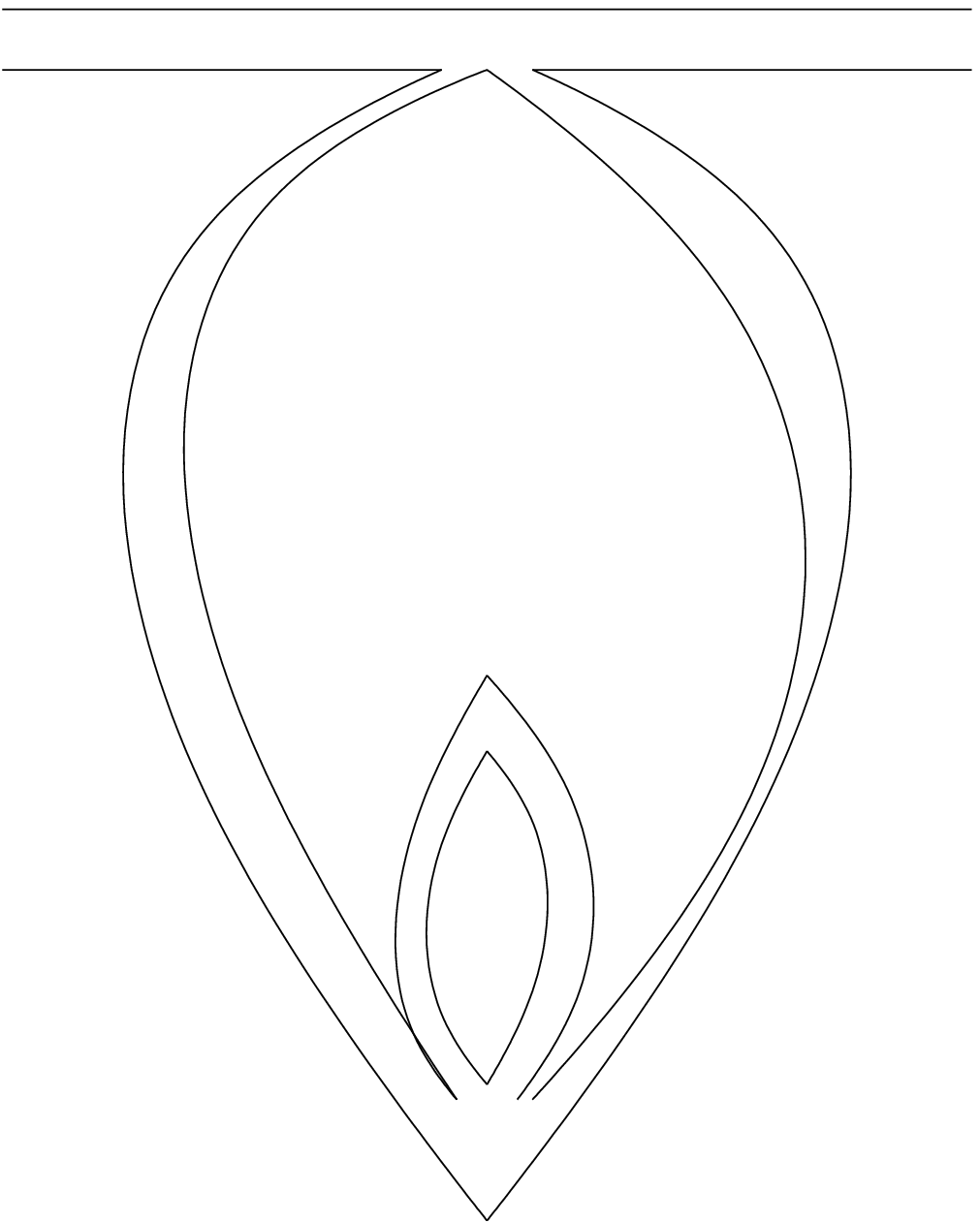}}}.
\eeqa
Applying the coproduct $\Delta$ on each of these ribbon graphs, one has (for the non-trivial part):
\beqa
\Delta' ({\includegraphics[scale=0.1]{propa-2l-1.eps}})&=&{\includegraphics[scale=0.05]{vertex-1l-1.eps}}\otimes ( {\includegraphics[scale=0.1]{tadpole-up.eps}}\, +
\, {\includegraphics[scale=0.1]{tadpole-down.eps}} ),\nonumber\\
\Delta' ({\includegraphics[scale=0.07]{propa-2l-2.eps}})&=&{\includegraphics[scale=0.05]{vertex-1l-2.eps}}\otimes  {\includegraphics[scale=0.1]{tadpole-up.eps}}
\, + \,
{\includegraphics[scale=0.1]{tadpole-up.eps}}\otimes  {\includegraphics[scale=0.1]{tadpole-up.eps}}
,\nonumber\\
\Delta' ({\includegraphics[scale=0.07]{propa-2l-3.eps}})&=&{\includegraphics[scale=0.1]{tadpole-down.eps}}\otimes  {\includegraphics[scale=0.1]{tadpole-up.eps}},\nonumber\\
\Delta' {\includegraphics[scale=0.07]{propa-2l-4.eps}}&=&{\includegraphics[scale=0.05]{vertex-1l-2.eps}}\otimes  {\includegraphics[scale=0.1]{tadpole-down.eps}}\, + \,
{\includegraphics[scale=0.1]{tadpole-down.eps}}\otimes  {\includegraphics[scale=0.1]{tadpole-down.eps}},\nonumber\\
\Delta' ({\includegraphics[scale=0.07]{propa-2l-5.eps}})&=&{\includegraphics[scale=0.1]{tadpole-up.eps}}\otimes  {\includegraphics[scale=0.1]{tadpole-down.eps}}.
\eeqa
Putting all this together leads to
\beqa
\label{rez1}
\Delta' (c_2^{{\includegraphics[scale=0.07]{propa.eps}}})= (c_1^{{\includegraphics[scale=0.07]{vertex.eps}}}+c_1^{{\includegraphics[scale=0.07]{propa.eps}}})\otimes c_1^{{\includegraphics[scale=0.07]{propa.eps}}}.
\eeqa

\bigskip

Let us now focus on the more involved case of the four-point function.  Using the definition \eqref{def-ckr}, one gets
\beqa
\label{c2v}
c_2^{{\includegraphics[scale=0.07]{vertex.eps}}}
&&=
{{\includegraphics[scale=0.15]{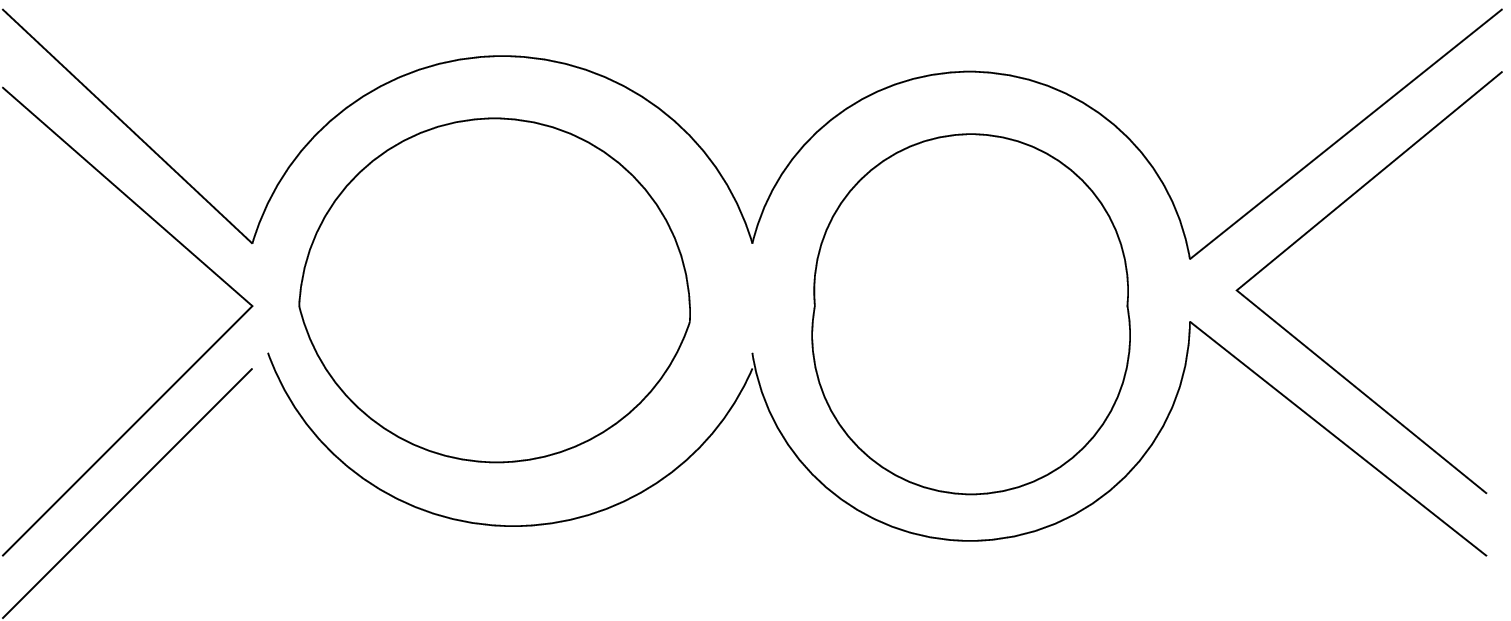}}}\, 
+ {{\includegraphics[scale=0.1]{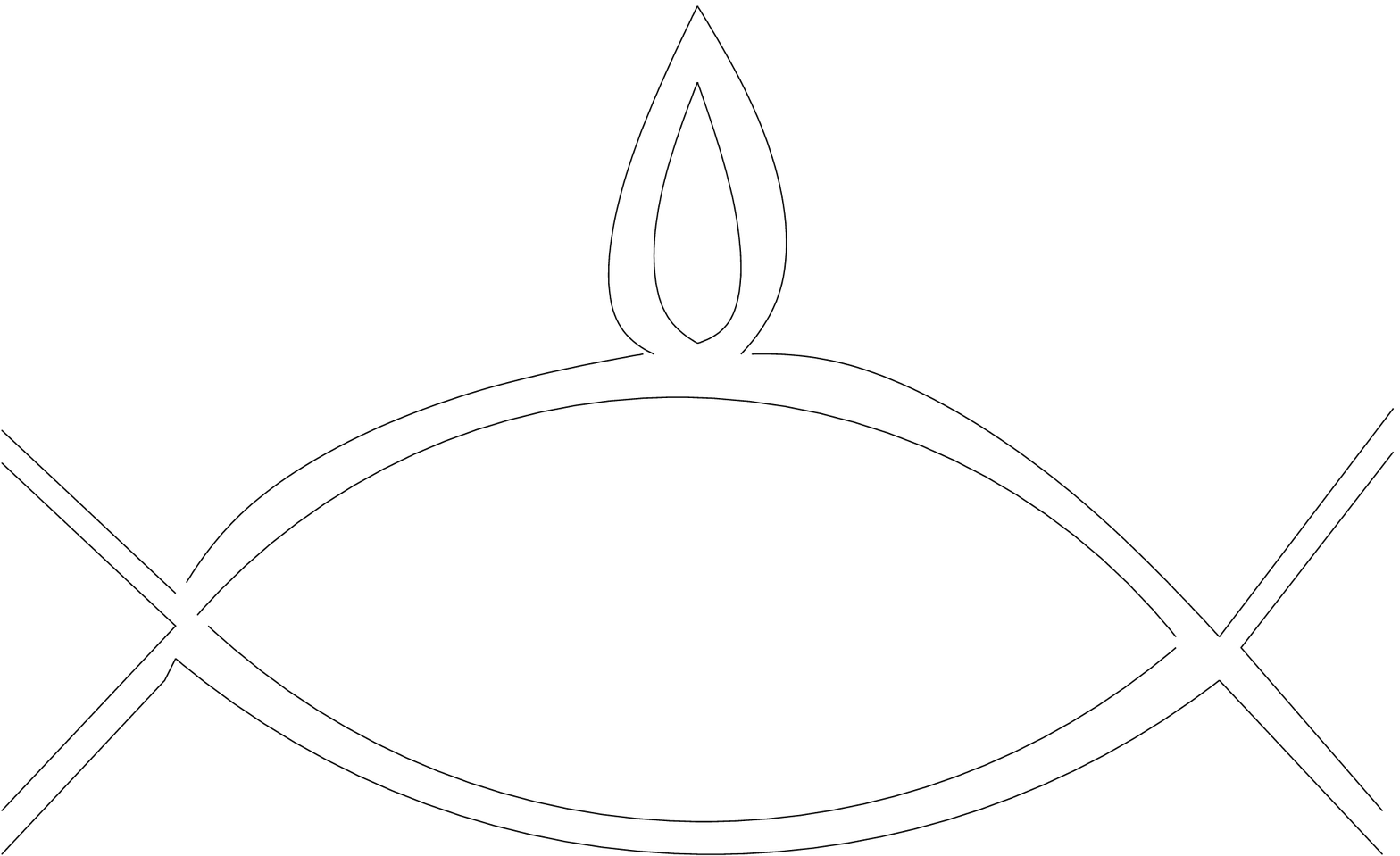}}}\, 
+ {{\includegraphics[scale=0.1]{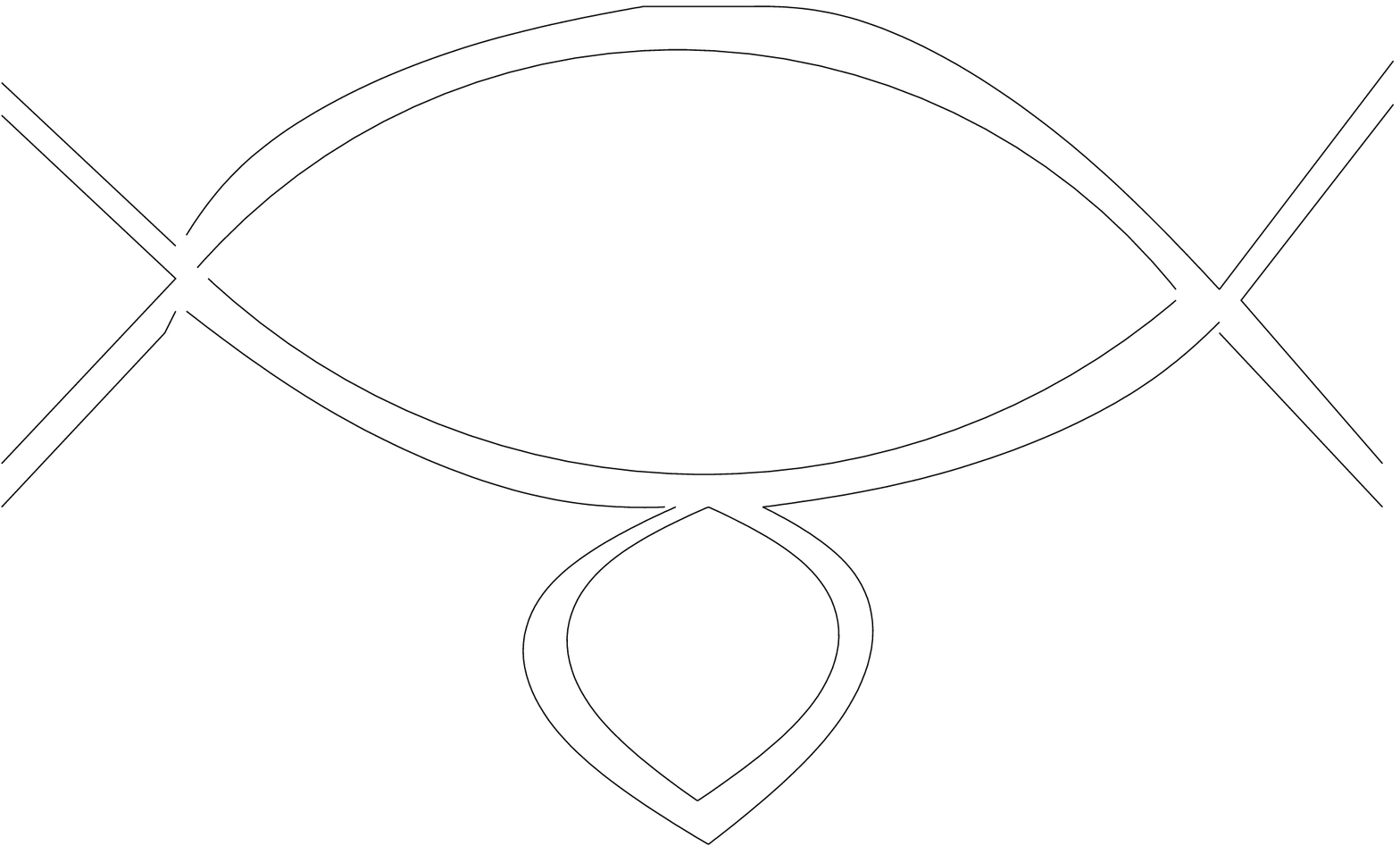}}}\, 
+ {{\includegraphics[scale=0.1]{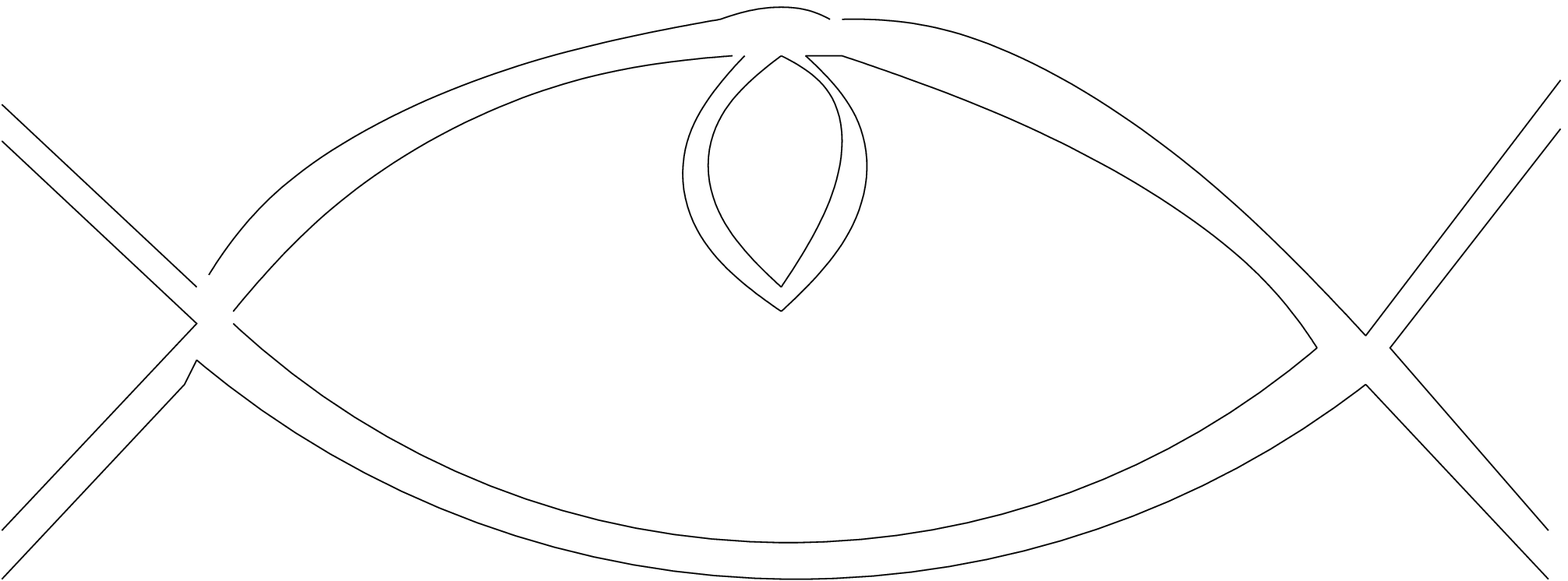}}}\, 
+ {{\includegraphics[scale=0.1]{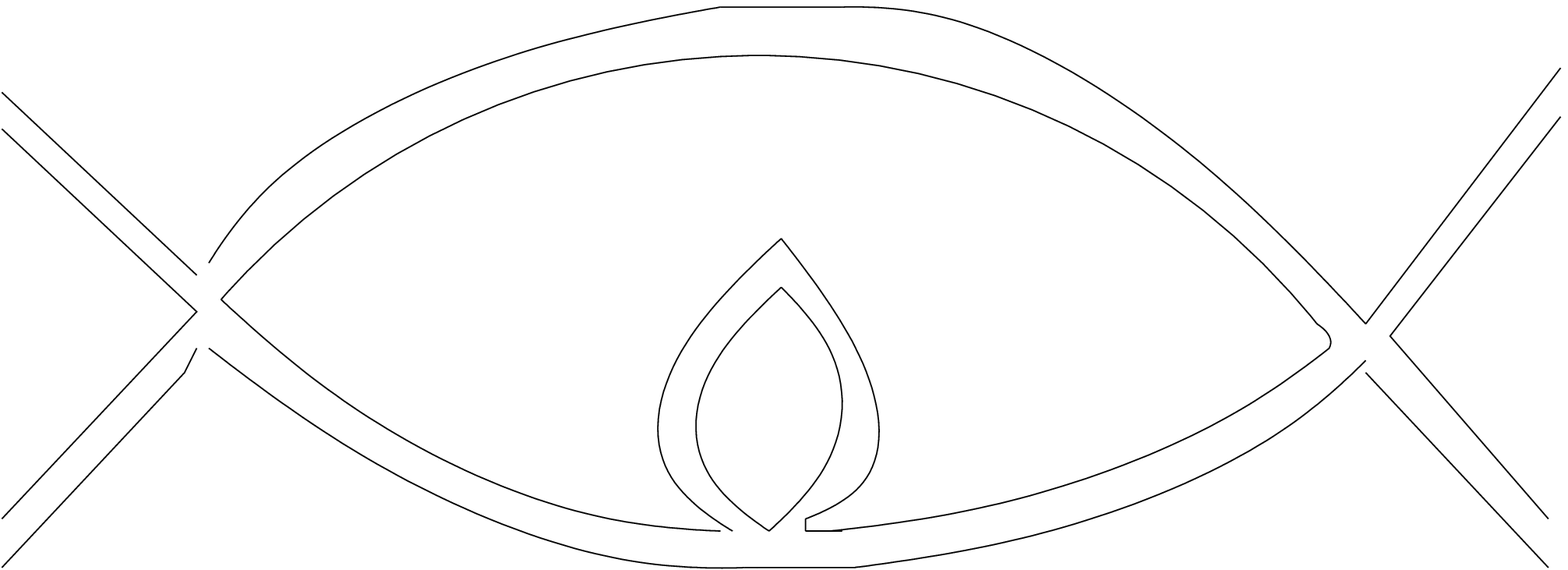}}}\nonumber\\
&&+{{\includegraphics[angle=90,scale=0.15]{vertex-2l-1.eps}}}\, 
+ {{\includegraphics[angle=90,scale=0.1]{vertex-2l-2.eps}}}\, 
+ {{\includegraphics[angle=90,scale=0.1]{vertex-2l-3.eps}}}\, 
+ {{\includegraphics[angle=90,scale=0.1]{vertex-2l-4.eps}}}\, 
+ {{\includegraphics[angle=90,scale=0.1]{vertex-2l-5.eps}}}\nonumber\\
&&+{{\includegraphics[scale=0.1]{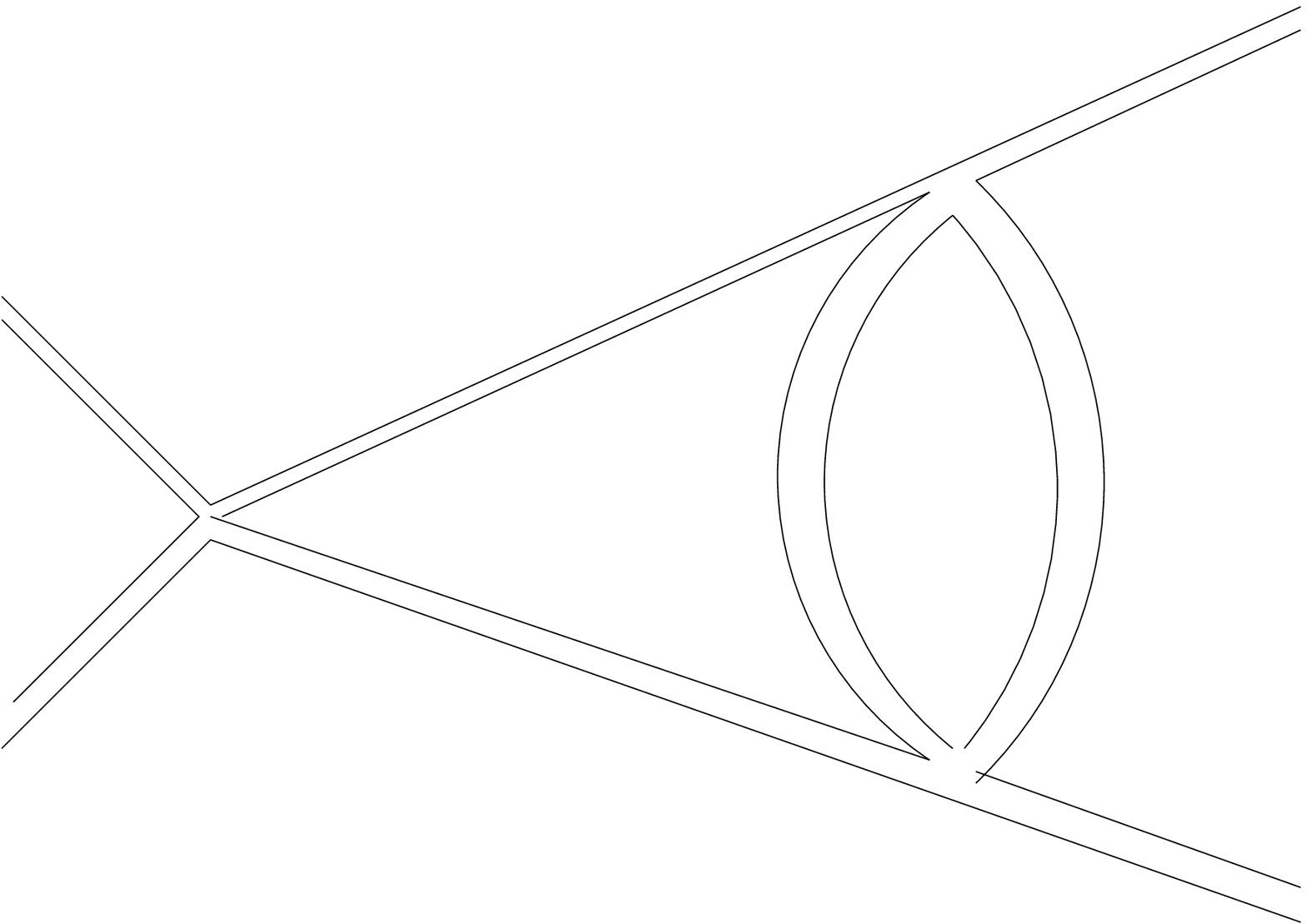}}}\,
+{{\includegraphics[angle=90,scale=0.1]{vertex-2l-11.eps}}}\,
+{{\includegraphics[angle=180,scale=0.1]{vertex-2l-11.eps}}}\,
+{{\includegraphics[angle=270,scale=0.1]{vertex-2l-11.eps}}}.
\eeqa
Applying the coproduct $\Delta$ on each of these ribbon graphs, one has (for the non-trivial part):
\beqa
\Delta' ({{\includegraphics[scale=0.1]{vertex-2l-1.eps}}})&=&2{{\includegraphics[scale=0.05]{vertex-1l-1.eps}}}\otimes {{\includegraphics[scale=0.05]{vertex-1l-1.eps}}},\nonumber\\
\Delta' ({{\includegraphics[scale=0.1]{vertex-2l-2.eps}}})&=&{{\includegraphics[scale=0.1]{tadpole-up.eps}}}\otimes {{\includegraphics[scale=0.05]{vertex-1l-1.eps}}},\nonumber\\
\Delta' ({{\includegraphics[angle=90,scale=0.1]{vertex-2l-1.eps}}})&=&2{{\includegraphics[scale=0.05]{vertex-1l-2.eps}}}\otimes {{\includegraphics[scale=0.05]{vertex-1l-2.eps}}},\nonumber\\
\Delta' ({{\includegraphics[scale=0.1]{vertex-2l-11.eps}}})&=&{{\includegraphics[scale=0.05]{vertex-1l-1.eps}}}\otimes {{\includegraphics[scale=0.05]{vertex-1l-2.eps}}}.
\eeqa
The remaining ten ribbon graphs on the RHS of \eqref{c2v} are treated analogously, finally leading to
\beqa
\label{rez2}
\Delta' (c_2^{{\includegraphics[scale=0.07]{vertex.eps}}})= (2c_1^{{\includegraphics[scale=0.07]{vertex.eps}}}+2c_1^{{\includegraphics[scale=0.07]{propa.eps}}})\otimes c_1^{{\includegraphics[scale=0.07]{vertex.eps}}}
\eeqa
(where we have used \eqref{c1uri}). The results \eqref{rez1} and \eqref{rez2} are thus illustrations of Theorem \ref{toate} {\it iii)}; these equations further give the expressions of the polynomials $P_1^{{\includegraphics[scale=0.07]{propa.eps}}}$ and $P_1^{{\includegraphics[scale=0.07]{vertex.eps}}}$.

\bigskip

Let us now show that each such two-loop graph lies in the image of our Hochschild one-cocycles  $B_+^{1,{\includegraphics[scale=0.07]{propa.eps}}}$. Using the definition \eqref{b+} and computing 
the combinatorial factors as indicated in section \ref{more}, 
one has
\beqa
\label{p1}
B_+^{{\includegraphics[scale=0.07]{tadpole-up.eps}}}({{\includegraphics[scale=0.05]{vertex-1l-1.eps}}}\ + \ {{\includegraphics[scale=0.05]{vertex-1l-2.eps}}})&=&
\frac 12 ({\includegraphics[scale=0.1]{propa-2l-1.eps}} \, +\,  {\includegraphics[scale=0.07]{propa-2l-2.eps}}),\nonumber\\
B_+^{{\includegraphics[scale=0.07]{tadpole-down.eps}}}({{\includegraphics[scale=0.05]{vertex-1l-1.eps}}}\ + \ {{\includegraphics[scale=0.05]{vertex-1l-2.eps}}})&=&
\frac 12 ({\includegraphics[scale=0.1]{propa-2l-1.eps}} \, +\,  {\includegraphics[scale=0.07]{propa-2l-4.eps}}),\nonumber\\
B_+^{{\includegraphics[scale=0.07]{tadpole-up.eps}}}({{\includegraphics[scale=0.1]{tadpole-up.eps}}}\ + \ {{\includegraphics[scale=0.1]{tadpole-down.eps}}})&=&
\frac 12 ({\includegraphics[scale=0.07]{propa-2l-2.eps}} \, +\,  {\includegraphics[scale=0.07]{propa-2l-3.eps}}),\nonumber\\
B_+^{{\includegraphics[scale=0.07]{tadpole-down.eps}}}({{\includegraphics[scale=0.1]{tadpole-up.eps}}}\ + \ {{\includegraphics[scale=0.1]{tadpole-down.eps}}})&=&
\frac 12 ({\includegraphics[scale=0.07]{propa-2l-5.eps}} \, +\,  {\includegraphics[scale=0.07]{propa-2l-4.eps}}).
\eeqa
The $\frac 12$ coefficients above come from the computation of the permutation of external legs, number of bijections and number of maximal forests for each of the resulting ribbon graphs, as explained in section \ref{more}. 
When adding up all this, one does not obtain $c_2^{{\includegraphics[scale=0.07]{propa.eps}}}$ (as given by \eqref{c2}). This comes from the fact that we have not yet included the planar irregular sector, which gives birth through these insertions to planar regular graphs (as explained in the subsection \ref{preLie}). One has
\beqa
\label{p2}
B_+^{{\includegraphics[scale=0.1]{tadpole-up.eps}}}( {\includegraphics[scale=0.1]{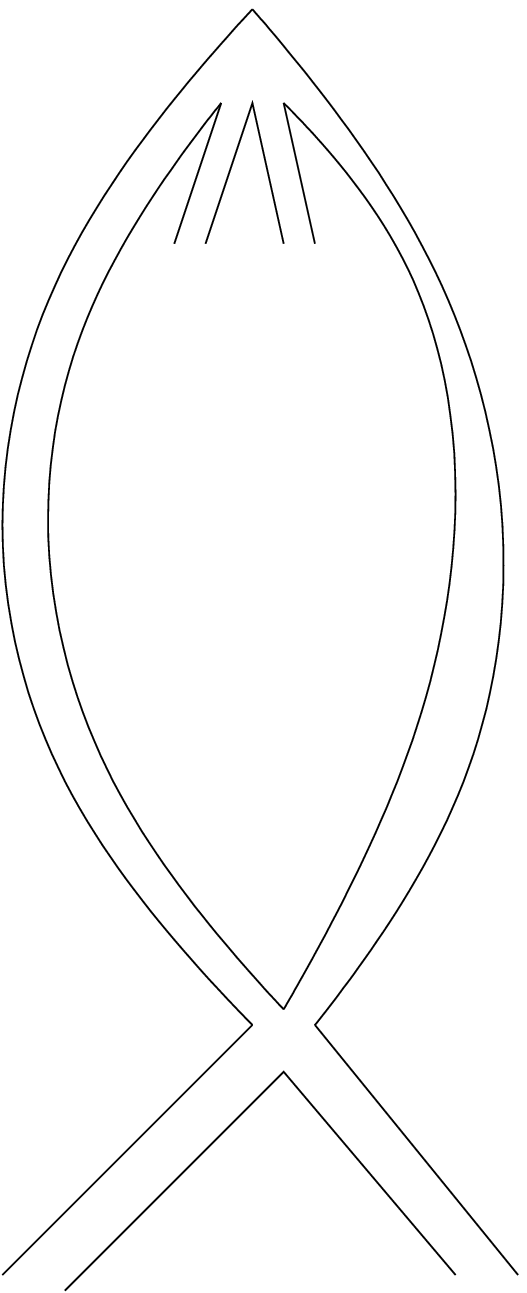}})&= &\frac 12 \, {\includegraphics[scale=0.15]{propa-2l-3.eps}},\nonumber\\
B_+^{{\includegraphics[scale=0.1]{tadpole-down.eps}}}( {\includegraphics[scale=0.1]{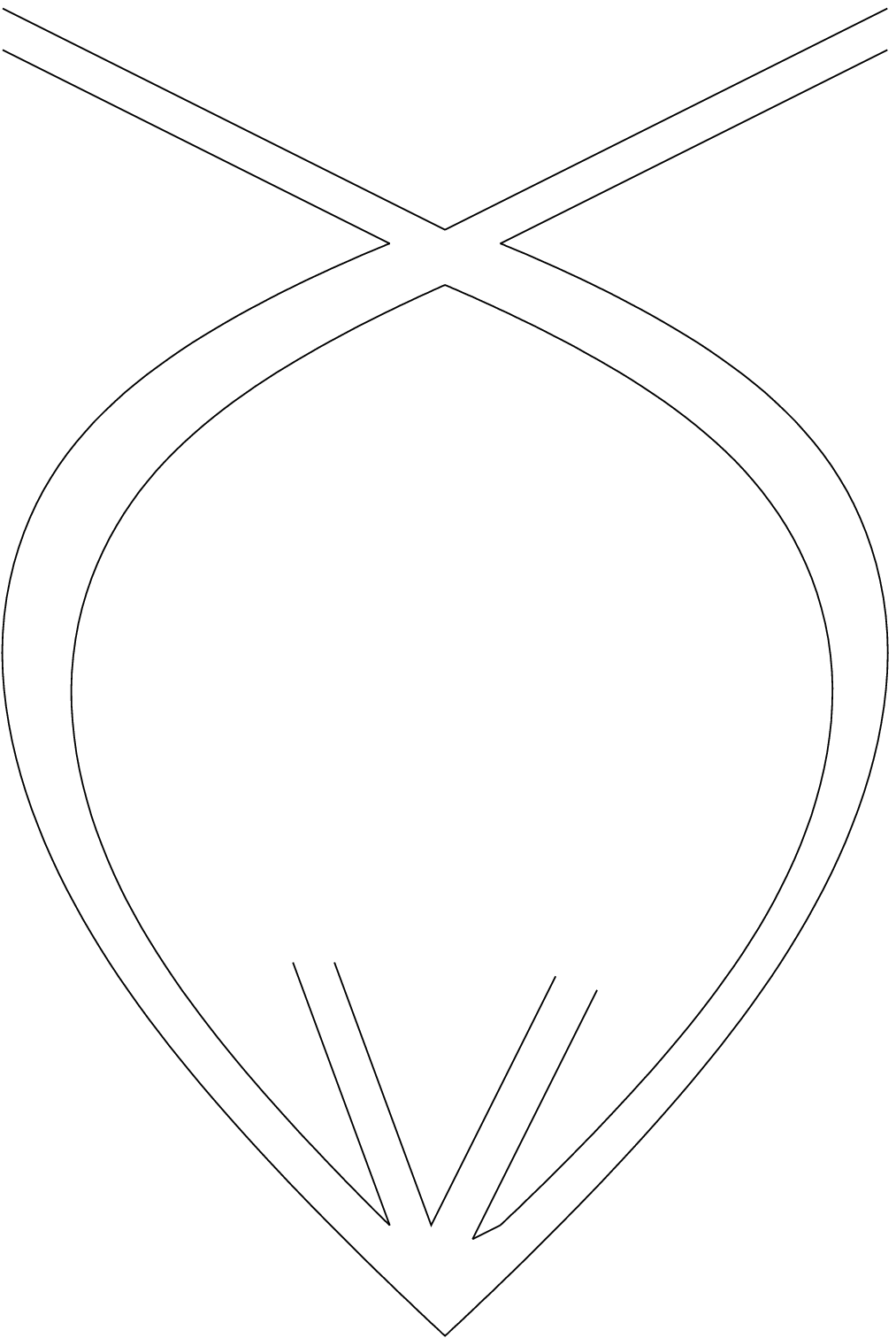}})&= & \frac 12\, {\includegraphics[scale=0.15]{propa-2l-5.eps}}.
\eeqa
The two new graphs above belong to 
\beqa
\label{tildec1p}
\widetilde c_1^{{\includegraphics[scale=0.07]{vertex.eps}}}
\eeqa
(which corresponds to the planar irregular sector). Note that the rest of the planar irregular graphs belonging to \eqref{tildec1p} do not lead to planar regular graphs when acting upon with  $B_+^{1,{\includegraphics[scale=0.07]{propa.eps}}}$.
Furthermore, one can analogously define
\beqa
\label{tildec1v}
\widetilde c_1^{{\includegraphics[scale=0.07]{propa.eps}}}= {\includegraphics[scale=0.15]{tadpole-np.eps}}.
\eeqa
Acting on this graph with  $B_+^{1,{\includegraphics[scale=0.07]{propa.eps}}}$ does not lead to a planar regular graph. Thus, adding up \eqref{p1} and \eqref{p2}, one obtains indeed $c_2^{{\includegraphics[scale=0.07]{propa.eps}}}$, as expected. We have thus proved that
\beqa
c_2^{{\includegraphics[scale=0.07]{propa.eps}}}=B_+^{1,{\includegraphics[scale=0.07]{propa.eps}}} (c_1^{{\includegraphics[scale=0.07]{propa.eps}}}+c_1^{{\includegraphics[scale=0.07]{vertex.eps}}}+ 
\widetilde c_1^{{\includegraphics[scale=0.07]{propa.eps}}}
+\widetilde c_1^{{\includegraphics[scale=0.07]{vertex.eps}}}).
\eeqa

\bigskip

Let us now explicitly show the necessity of adding the irregular sector also when writing down Theorem \ref{toate} {\it ii)} at this two loop level. 
In order to do this we first consider the four planar regular graphs:
\beqa
\label{simplele}
\includegraphics[scale=0.15]{tadpole-up.eps},
\includegraphics[scale=0.07]{tadpole-down.eps}, 
\includegraphics[scale=0.07]{vertex-1l-1.eps},
\includegraphics[scale=0.07]{vertex-1l-2.eps}.
\eeqa
Using \eqref{p1} and the definition \eqref{coproduct} of the coproduct one can write down the LHS and the RHS of  Theorem \ref{toate} {\it ii)}. Let us right down the contribution of the graphs of \eqref{simplele} to the LHS of  Theorem \ref{toate} {\it ii)}.
One has the following relations:
\begin{enumerate}
\item
\beqa
\label{l1}
\Delta (B_+^{1,{\includegraphics[scale=0.07]{propa.eps}}}(\includegraphics[scale=0.1]{tadpole-up.eps}))=\Delta((B_+^{{\includegraphics[scale=0.07]{tadpole-up.eps}}}+B_+^{{\includegraphics[scale=0.07]{tadpole-down.eps}}})(\includegraphics[scale=0.1]{tadpole-up.eps})))=
\frac 12 \Delta({\includegraphics[scale=0.07]{propa-2l-2.eps}}+{\includegraphics[scale=0.07]{propa-2l-5.eps}})\nonumber\\
=\frac 12 {\includegraphics[scale=0.07]{propa-2l-2.eps}}\otimes 1_\ch+
\frac 12 1_\ch\otimes {\includegraphics[scale=0.07]{propa-2l-2.eps}}
+\frac 12  {{\includegraphics[scale=0.05]{vertex-1l-2.eps}}}\otimes \includegraphics[scale=0.1]{tadpole-up.eps}+\frac 12 \includegraphics[scale=0.1]{tadpole-up.eps}\otimes \includegraphics[scale=0.1]{tadpole-up.eps}\nonumber\\
+ \frac 12 {\includegraphics[scale=0.07]{propa-2l-5.eps}}\otimes 1_\ch+
\frac 12 1_\ch\otimes {\includegraphics[scale=0.07]{propa-2l-5.eps}}
+ \frac 12 \includegraphics[scale=0.1]{tadpole-up.eps}\otimes \includegraphics[scale=0.1]{tadpole-down.eps}.
\eeqa

\item

\beqa
\label{l2}
\Delta (B_+^{1,{\includegraphics[scale=0.07]{propa.eps}}}(\includegraphics[scale=0.07]{tadpole-down.eps}))=\Delta((B_+^{{\includegraphics[scale=0.07]{tadpole-up.eps}}}+B_+^{{\includegraphics[scale=0.07]{tadpole-down.eps}}})(\includegraphics[scale=0.07]{tadpole-down.eps})))=
\frac 12 \Delta({\includegraphics[scale=0.07]{propa-2l-3.eps}}+{\includegraphics[scale=0.07]{propa-2l-4.eps}})\nonumber\\
=\frac 12 {\includegraphics[scale=0.07]{propa-2l-3.eps}}\otimes 1_\ch+
\frac 12 1_\ch\otimes {\includegraphics[scale=0.07]{propa-2l-3.eps}}
+\frac 12 \includegraphics[scale=0.07]{tadpole-down.eps}\otimes \includegraphics[scale=0.1]{tadpole-up.eps}\nonumber\\
+ \frac 12 {\includegraphics[scale=0.07]{propa-2l-4.eps}}\otimes 1_\ch+
\frac 12 1_\ch\otimes {\includegraphics[scale=0.07]{propa-2l-4.eps}}
+\frac 12  {{\includegraphics[scale=0.05]{vertex-1l-2.eps}}}\otimes \includegraphics[scale=0.07]{tadpole-down.eps}
+ \frac 12 \includegraphics[scale=0.07]{tadpole-down.eps}\otimes \includegraphics[scale=0.07]{tadpole-down.eps}.
\eeqa

\item 

\beqa
\label{l3}
\Delta (B_+^{1,{\includegraphics[scale=0.07]{propa.eps}}}(\includegraphics[scale=0.07]{vertex-1l-1.eps}))=\Delta((B_+^{{\includegraphics[scale=0.07]{tadpole-up.eps}}}+B_+^{{\includegraphics[scale=0.07]{tadpole-down.eps}}})(\includegraphics[scale=0.07]{vertex-1l-1.eps})))=
\Delta({\includegraphics[scale=0.15]{propa-2l-1.eps}})\nonumber\\
= {\includegraphics[scale=0.15]{propa-2l-1.eps}}\otimes 1_\ch+
\ 1_\ch\otimes {\includegraphics[scale=0.15]{propa-2l-1.eps}}
+  \frac 12 {{\includegraphics[scale=0.05]{vertex-1l-1.eps}}}\otimes \includegraphics[scale=0.1]{tadpole-up.eps}+
+ \frac 12 \includegraphics[scale=0.05]{vertex-1l-1.eps}\otimes \includegraphics[scale=0.07]{tadpole-down.eps}.
\eeqa

\item

\beqa
\label{l4}
\Delta (B_+^{1,{\includegraphics[scale=0.07]{propa.eps}}}(\includegraphics[scale=0.07]{vertex-1l-2.eps}))=\Delta((B_+^{{\includegraphics[scale=0.07]{tadpole-up.eps}}}+B_+^{{\includegraphics[scale=0.07]{tadpole-down.eps}}})(\includegraphics[scale=0.07]{vertex-1l-2.eps})))=
\frac 12 \Delta({\includegraphics[scale=0.1]{propa-2l-2.eps}}
+{\includegraphics[scale=0.1]{propa-2l-4.eps}})\nonumber\\
= \frac 12 {\includegraphics[scale=0.1]{propa-2l-2.eps}}\otimes 1_\ch+
\frac 12 1_\ch\otimes {\includegraphics[scale=0.1]{propa-2l-2.eps}}
+  {{\includegraphics[scale=0.1]{vertex-1l-2.eps}}}\otimes \includegraphics[scale=0.1]{tadpole-up.eps}+
+ \frac 12 \includegraphics[scale=0.1]{tadpole-up.eps}\otimes \includegraphics[scale=0.07]{tadpole-down.eps}\nonumber\\
\frac 12 {\includegraphics[scale=0.1]{propa-2l-4.eps}}\otimes 1_\ch+
\frac 12 1_\ch\otimes {\includegraphics[scale=0.1]{propa-2l-4.eps}}
+  {{\includegraphics[scale=0.1]{vertex-1l-2.eps}}}\otimes \includegraphics[scale=0.07]{tadpole-down.eps}+
+ \frac 12 \includegraphics[scale=0.07]{tadpole-down.eps}\otimes \includegraphics[scale=0.07]{tadpole-down.eps}.
\eeqa

\end{enumerate}

Adding together the contributions of equations \eqref{l1} to \eqref{l4} one has for the LHS of Theorem \ref{toate} {\it ii)} the following results (corresponding to the planar regular graphs listed in equation \eqref{simplele}):
\beqa
\label{l-tot}
\frac 12 {\includegraphics[scale=0.07]{propa-2l-2.eps}}\otimes 1_\ch+
\frac 12 1_\ch\otimes {\includegraphics[scale=0.07]{propa-2l-2.eps}}
+\frac 12  {{\includegraphics[scale=0.05]{vertex-1l-2.eps}}}\otimes \includegraphics[scale=0.1]{tadpole-up.eps}+\frac 12 \includegraphics[scale=0.1]{tadpole-up.eps}\otimes \includegraphics[scale=0.1]{tadpole-up.eps}\nonumber\\
+ \frac 12 {\includegraphics[scale=0.07]{propa-2l-5.eps}}\otimes 1_\ch+
\frac 12 1_\ch\otimes {\includegraphics[scale=0.07]{propa-2l-5.eps}}
+ \frac 12 \includegraphics[scale=0.1]{tadpole-up.eps}\otimes \includegraphics[scale=0.1]{tadpole-down.eps}\nonumber\\
+\frac 12 {\includegraphics[scale=0.07]{propa-2l-3.eps}}\otimes 1_\ch+
\frac 12 1_\ch\otimes {\includegraphics[scale=0.07]{propa-2l-3.eps}}
+\frac 12 \includegraphics[scale=0.07]{tadpole-down.eps}\otimes \includegraphics[scale=0.1]{tadpole-up.eps}\nonumber\\
+ \frac 12 {\includegraphics[scale=0.07]{propa-2l-4.eps}}\otimes 1_\ch+
\frac 12 1_\ch\otimes {\includegraphics[scale=0.07]{propa-2l-4.eps}}
+\frac 12  {{\includegraphics[scale=0.05]{vertex-1l-2.eps}}}\otimes \includegraphics[scale=0.07]{tadpole-down.eps}
+ \frac 12 \includegraphics[scale=0.07]{tadpole-down.eps}\otimes \includegraphics[scale=0.07]{tadpole-down.eps}\nonumber\\
 + {\includegraphics[scale=0.15]{propa-2l-1.eps}}\otimes 1_\ch+
\ 1_\ch\otimes {\includegraphics[scale=0.15]{propa-2l-1.eps}}
+  \frac 12 {{\includegraphics[scale=0.05]{vertex-1l-1.eps}}}\otimes \includegraphics[scale=0.1]{tadpole-up.eps}+
+ \frac 12 \includegraphics[scale=0.05]{vertex-1l-1.eps}\otimes \includegraphics[scale=0.07]{tadpole-down.eps}\nonumber\\
+ \frac 12 {\includegraphics[scale=0.1]{propa-2l-2.eps}}\otimes 1_\ch+
\frac 12 1_\ch\otimes {\includegraphics[scale=0.1]{propa-2l-2.eps}}
+  {{\includegraphics[scale=0.1]{vertex-1l-2.eps}}}\otimes \includegraphics[scale=0.1]{tadpole-up.eps}+
+ \frac 12 \includegraphics[scale=0.1]{tadpole-up.eps}\otimes \includegraphics[scale=0.07]{tadpole-down.eps}\nonumber\\
+ \frac 12 {\includegraphics[scale=0.1]{propa-2l-4.eps}}\otimes 1_\ch+
\frac 12 1_\ch\otimes {\includegraphics[scale=0.1]{propa-2l-4.eps}}
+  {{\includegraphics[scale=0.1]{vertex-1l-2.eps}}}\otimes \includegraphics[scale=0.07]{tadpole-down.eps}+
+ \frac 12 \includegraphics[scale=0.07]{tadpole-down.eps}\otimes \includegraphics[scale=0.07]{tadpole-down.eps}.
\eeqa

The RHS of  Theorem \ref{toate} {\it ii)} corresponding to the total contribution of the planar regular sector listed in \eqref{simplele} is worked out analogously, leading to
\beqa
\label{r-tot}
\frac 12 {\includegraphics[scale=0.07]{propa-2l-2.eps}}\otimes 1_\ch+
\frac 12 1_\ch\otimes {\includegraphics[scale=0.07]{propa-2l-2.eps}}
+\frac 12  {{\includegraphics[scale=0.05]{vertex-1l-2.eps}}}\otimes \includegraphics[scale=0.1]{tadpole-up.eps}+\frac 12 \includegraphics[scale=0.1]{tadpole-up.eps}\otimes \includegraphics[scale=0.1]{tadpole-up.eps}\nonumber\\
+ \frac 12 {\includegraphics[scale=0.07]{propa-2l-5.eps}}\otimes 1_\ch+
\frac 12 1_\ch\otimes {\includegraphics[scale=0.07]{propa-2l-5.eps}}
+ \frac 12 \includegraphics[scale=0.1]{tadpole-up.eps}\otimes \includegraphics[scale=0.1]{tadpole-down.eps}\nonumber\\
+\frac 12 {\includegraphics[scale=0.07]{propa-2l-3.eps}}\otimes 1_\ch+
\frac 12 1_\ch\otimes {\includegraphics[scale=0.07]{propa-2l-3.eps}}
+ \includegraphics[scale=0.07]{tadpole-down.eps}\otimes \includegraphics[scale=0.1]{tadpole-up.eps}\nonumber\\
+ \frac 12 {\includegraphics[scale=0.07]{propa-2l-4.eps}}\otimes 1_\ch+
\frac 12 1_\ch\otimes {\includegraphics[scale=0.07]{propa-2l-4.eps}}
+\frac 12  {{\includegraphics[scale=0.05]{vertex-1l-2.eps}}}\otimes \includegraphics[scale=0.07]{tadpole-down.eps}
+ \frac 12 \includegraphics[scale=0.07]{tadpole-down.eps}\otimes \includegraphics[scale=0.07]{tadpole-down.eps}\nonumber\\
 +{\includegraphics[scale=0.15]{propa-2l-1.eps}}\otimes 1_\ch+
\ 1_\ch\otimes {\includegraphics[scale=0.15]{propa-2l-1.eps}}
+  \frac 12 {{\includegraphics[scale=0.05]{vertex-1l-1.eps}}}\otimes \includegraphics[scale=0.1]{tadpole-up.eps}+
+ \frac 12 \includegraphics[scale=0.05]{vertex-1l-1.eps}\otimes \includegraphics[scale=0.07]{tadpole-down.eps}\nonumber\\
 +\frac 12 {\includegraphics[scale=0.1]{propa-2l-2.eps}}\otimes 1_\ch+
\frac 12 1_\ch\otimes {\includegraphics[scale=0.1]{propa-2l-2.eps}}
+  {{\includegraphics[scale=0.1]{vertex-1l-2.eps}}}\otimes \includegraphics[scale=0.1]{tadpole-up.eps}+
+ \includegraphics[scale=0.1]{tadpole-up.eps}\otimes \includegraphics[scale=0.07]{tadpole-down.eps}\nonumber\\
+\frac 12 {\includegraphics[scale=0.1]{propa-2l-4.eps}}\otimes 1_\ch+
\frac 12 1_\ch\otimes {\includegraphics[scale=0.1]{propa-2l-4.eps}}
+  {{\includegraphics[scale=0.1]{vertex-1l-2.eps}}}\otimes \includegraphics[scale=0.07]{tadpole-down.eps}+
+ \frac 12 \includegraphics[scale=0.07]{tadpole-down.eps}\otimes \includegraphics[scale=0.07]{tadpole-down.eps}.
\eeqa
Comparing equations \eqref{l-tot} and \eqref{r-tot} above one is left on the RHS with
\beqa
\label{rest}
\frac 12 (\includegraphics[scale=0.15]{tadpole-up.eps} \otimes \includegraphics[scale=0.07]{tadpole-down.eps} + \includegraphics[scale=0.07]{tadpole-down.eps}\otimes \includegraphics[scale=0.15]{tadpole-up.eps}).
\eeqa
As above, the planar irregular sector  $\widetilde c_1^{{\includegraphics[scale=0.07]{vertex.eps}}}$ saves the day. Indeed, when computing the LHS contribution associated to this new sector one has (using \eqref{p2})
\beqa
\label{e1}
\Delta B_+^{1,{\includegraphics[scale=0.07]{propa.eps}}}(\widetilde c_1^{{\includegraphics[scale=0.07]{vertex.eps}}})=\frac 12 \Delta (  {\includegraphics[scale=0.15]{propa-2l-3.eps}}+
 {\includegraphics[scale=0.15]{propa-2l-5.eps}}).
\eeqa
Using again the definition \eqref{coproduct} of the coproduct, \eqref{e1} leads to
\beqa
\label{lhs}
&&\frac 12 (  {\includegraphics[scale=0.15]{propa-2l-3.eps}}\otimes 1_\ch + 1 \otimes  {\includegraphics[scale=0.15]{propa-2l-3.eps}}+
 {\includegraphics[scale=0.15]{propa-2l-5.eps}}\otimes 1_\ch + 1 \otimes  {\includegraphics[scale=0.15]{propa-2l-5.eps}}
\nonumber\\
&& 
+ \includegraphics[scale=0.15]{tadpole-up.eps} \otimes \includegraphics[scale=0.07]{tadpole-down.eps} + \includegraphics[scale=0.07]{tadpole-down.eps}\otimes \includegraphics[scale=0.15]{tadpole-up.eps}).
\eeqa
Let us now explicitly calculate the contribution of the new planar irregular sector  $\widetilde c_1^{{\includegraphics[scale=0.07]{vertex.eps}}}$ to the RHS of   Theorem \ref{toate} {\it ii)}. Using again \eqref{p2} and discarding the planar irregular graphs from the final list, one gets
\beqa
\label{rhs}
\frac 12 (  {\includegraphics[scale=0.15]{propa-2l-3.eps}}\otimes 1_\ch + 1 \otimes  {\includegraphics[scale=0.15]{propa-2l-3.eps}}+
 {\includegraphics[scale=0.15]{propa-2l-5.eps}}\otimes 1_\ch + 1 \otimes  {\includegraphics[scale=0.15]{propa-2l-5.eps}}.
\eeqa
This cancels out with the LHS contribution of \eqref{lhs}. One can thus see that the planar irregular sector has finally led to a total contribution 
\beqa
\frac 12 (\includegraphics[scale=0.15]{tadpole-up.eps} \otimes \includegraphics[scale=0.07]{tadpole-down.eps} + \includegraphics[scale=0.07]{tadpole-down.eps}\otimes \includegraphics[scale=0.15]{tadpole-up.eps})
\eeqa
in the LHS of  Theorem \ref{toate} {\it ii)}. This cancels out with the rest \eqref{rest} of the planar regular sector. Let us remark that taking into consideration the planar irregular tadpole \eqref{tildec1v} does not change the situation since the insertion of this graph leads directly to non-planar graphs which are to be discarded. We have thus completely checked out  Theorem \ref{toate} {\it ii)} at the two-loop level, as announced. 

\bigskip

{\bf Acknowledgment:} A. T. acknowledges J. Magnen and T. Krajewski for discussions. D. K. thanks K. Yeats for discussions. A. T. was partially supported by CNCSIS grant ``Idei'' ID-44, 454/2009. D.K. was partially supported by NSF grant DMS-0603781

\end{document}